\documentclass[12pt,a4paper]{article}
\usepackage{amsmath,amssymb,graphicx,epsfig,cite,color}
\usepackage[left=2.7cm,right=2.7cm,top=2.5cm,bottom=2.5cm]{geometry}
\usepackage{subfigure}
\usepackage[table]{xcolor}
\usepackage{booktabs}

\usepackage{graphicx}
\usepackage{lscape}
\usepackage{appendix}
\usepackage{multirow}
\usepackage[colorlinks=true,linkcolor=red]{hyperref}%

\begin{document}

\def\ds{\displaystyle}
\def\beq{\begin{equation}}
\def\eeq{\end{equation}}
\def\bea{\begin{eqnarray}}
\def\eea{\end{eqnarray}}
\def\beeq{\begin{eqnarray}}
\def\eeeq{\end{eqnarray}}

\def\rar{\rightarrow} 
\def\nnb{\nonumber}

\def\ds{\displaystyle}
\def\beq{\begin{equation}}
\def\eeq{\end{equation}}
\def\bea{\begin{eqnarray}}
\def\eea{\end{eqnarray}}
\def\beeq{\begin{eqnarray}}
\def\eeeq{\end{eqnarray}}
\def\ve{\vert}
\def\vel{\left|}
\def\ver{\right|}
\def\nnb{\nonumber}
\def\ga{\left(}
\def\dr{\right)}
\def\aga{\left\{}
\def\adr{\right\}}
\def\lla{\left<}
\def\rra{\right>}
\def\rar{\rightarrow}
\def\lrar{\leftrightarrow}
\def\nnb{\nonumber}
\def\la{\langle}
\def\ra{\rangle}
\def\ba{\begin{array}}
\def\ea{\end{array}}
\def\tr{\mbox{Tr}}
\def\ssp{{\Sigma^{*+}}}
\def\sso{{\Sigma^{*0}}}
\def\ssm{{\Sigma^{*-}}}
\def\xis0{{\Xi^{*0}}}
\def\xism{{\Xi^{*-}}}
\def\qs{\la \bar s s \ra}
\def\qu{\la \bar u u \ra}
\def\qd{\la \bar d d \ra}
\def\qq{\la \bar q q \ra}
\def\gGgG{\la g^2 G^2 \ra}
\def\q{\gamma_5 \not\!q}
\def\x{\gamma_5 \not\!x}
\def\g5{\gamma_5}
\def\sb{S_Q^{cf}}
\def\sd{S_d^{be}}
\def\su{S_u^{ad}}
\def\sbp{{S}_Q^{'cf}}
\def\sdp{{S}_d^{'be}}
\def\sup{{S}_u^{'ad}}
\def\ssp{{S}_s^{'??}}

\def\sig{\sigma_{\mu \nu} \gamma_5 p^\mu q^\nu}
\def\fo{f_0(\frac{s_0}{M^2})}
\def\ffi{f_1(\frac{s_0}{M^2})}
\def\fii{f_2(\frac{s_0}{M^2})}
\def\O{{\cal O}}
\def\sl{{\Sigma^0 \Lambda}}
\def\es{\!\!\! &=& \!\!\!}
\def\ap{\!\!\! &\approx& \!\!\!}
\def\ar{&+& \!\!\!}
\def\ek{&-& \!\!\!}
\def\kek{\!\!\!&-& \!\!\!}
\def\cp{&\times& \!\!\!}
\def\se{\!\!\! &\simeq& \!\!\!}
\def\eqv{&\equiv& \!\!\!}
\def\kpm{&\pm& \!\!\!}
\def\kmp{&\mp& \!\!\!}
\def\mcdot{\!\cdot\!}
\def\erar{&\rightarrow&}



\renewcommand{\textfraction}{0.2}    
\renewcommand{\topfraction}{0.8}   

\renewcommand{\bottomfraction}{0.4}   
\renewcommand{\floatpagefraction}{0.8}
\newcommand\mysection{\setcounter{equation}{0}\section}

\def\baeq{\begin{appeq}}     \def\eaeq{\end{appeq}}  
\def\baeeq{\begin{appeeq}}   \def\eaeeq{\end{appeeq}}
\newenvironment{appeq}{\beq}{\eeq}   
\newenvironment{appeeq}{\beeq}{\eeeq}
\def\bAPP#1#2{
 \markright{APPENDIX #1}
 \addcontentsline{toc}{section}{Appendix #1: #2}
 \medskip
 \medskip
 \begin{center}      {\bf\LARGE Appendix #1 :}{\quad\Large\bf #2}
\end{center}
 \renewcommand{\thesection}{#1.\arabic{section}}
\setcounter{equation}{0}
        \renewcommand{\thehran}{#1.\arabic{hran}}
\renewenvironment{appeq}
  {  \renewcommand{\theequation}{#1.\arabic{equation}}
     \beq
  }{\eeq}
\renewenvironment{appeeq}
  {  \renewcommand{\theequation}{#1.\arabic{equation}}
     \beeq
  }{\eeeq}
\nopagebreak \noindent}

\def\eAPP{\renewcommand{\thehran}{\thesection.\arabic{hran}}}

\renewcommand{\theequation}{\arabic{equation}}
\newcounter{hran}
\renewcommand{\thehran}{\thesection.\arabic{hran}}

\def\bmini{\setcounter{hran}{\value{equation}}
\refstepcounter{hran}\setcounter{equation}{0}
\renewcommand{\theequation}{\thehran\alph{equation}}\begin{eqnarray}}
\def\bminiG#1{\setcounter{hran}{\value{equation}}
\refstepcounter{hran}\setcounter{equation}{-1}
\renewcommand{\theequation}{\thehran\alph{equation}}
\refstepcounter{equation}\label{#1}\begin{eqnarray}}


\newskip\humongous \humongous=0pt plus 1000pt minus 1000pt
\def\caja{\mathsurround=0pt}
 

\title{
         {\Large
                 {\bf
Strong Coupling Constants of the Doubly Heavy Spin-1/2 Baryons with Light Pseudoscalar Mesons  
                 }
         }
      }

\author{\vspace{1cm}\\
{\small
S.~Rostami$^{1}$,
K. Azizi$^{1,2,4}$,
A. R. Olamaei$^{3,4}$} \\
{\small $^1$ Department of Physics, University of Tehran, North Karegar Ave. Tehran 14395-547, Iran}\\
{\small $^2$ Department of Physics, Do\v{g}u\c{s} University, Acibadem-Kadik\"{o}y, 34722
Istanbul, Turkey}\\
{\small $^3$ Department of Physics, Jahrom University, Jahrom, P.~ O.~ Box 74137-66171, Iran}\\
{\small$^4$  School of Particles and Accelerators, Institute for Research in Fundamental Sciences (IPM), } \\
 {\small P. O. Box 19395-5531, Tehran, Iran }}

\date{}

\begin{titlepage}
\maketitle
\thispagestyle{empty}

\begin{abstract}
The strong coupling constants of hadronic multiplets are fundamental parameters which carry information about the strong interactions among participating particles. These parameters can help us construct the hadron-hadron strong potential and gain information about the structure of the involved hadrons. Motivated by the recent observation of the doubly charmed $\Xi_{cc}$ state by LHCb, we determine  the strong coupling constants  among the doubly heavy spin-1/2 baryons, $  \Xi^{(\prime)}_{QQ^\prime }$, $ \Omega^{(\prime)}_{QQ^\prime}$ and light pseudoscalar mesons, $ \pi $, $ K $,  $\eta$ and $ \eta^\prime $ within the framework of the light cone QCD sum rules. The obtained results may help experimental groups in  analysis of the related data at hadron colliders. 

\end{abstract}

\end{titlepage}
\section{Introduction}

The quark model \cite{GELLMANN1964214,Zweig:1981pd,Zweig:1964jf}  has been very successful in describing the properties of 
 hadrons observed  experimentally. Nevertheless, 
not all particles predicted by the quark model are  experimentally well established: of the doubly
heavy baryons, only  the doubly charmed $\Xi_{cc}$ state has been seen in  experiment. The triply heavy  baryons are also missing in  experiments and  hunt for them   continues. More experimental and theoretical studies on these states are required. Even in the case of  $\Xi_{cc}$ there is a  puzzle  in the experimental results. The
first evidence for this state was reported in 2005 by the SELEX experiment, with   
 $\Xi_{cc}^{+}$ decaying into $ \Lambda_c^+ K^- \pi^+ $ and $pD^+K^-  $   final states,
using a $600\text{MeV}/c^2$ charged hyperon beam impinging on a fixed target.
The mass measured by SELEX, averaged over the two decay
modes, was found to be $ 3518.7\pm 1.7~\text{MeV}/c^2 $.
The lifetime was measured to be less than $ 33~\text{fs} $  at $ 90\% $ confidence level.
It was estimated that about $  20\% $ of $ \Lambda_c^+ $ baryons in the SELEX experiment were produced from  $\Xi_{cc}$  decays~\cite{Mattson:2002vu,Ocherashvili:2004hi}.
However, the FOCUS \cite{Ratti:2003ez}, BaBar \cite{Aubert:2006qw}, LHCb \cite{Aaij:2013voa} and Belle \cite{Kato:2013ynr} experiments
were not able to confirm the SELEX results. 
In 2017, the doubly charmed baryon $ \Xi^{++}_{cc} $ was
observed by the LHCb collaboration via the decay channel, $\Xi^{++}_{cc}\rightarrow\Lambda_c^+ K^-\pi^+\pi^+$
\cite{Aaij:2017ueg}, 
and confirmed via measuring another decay channel $\Xi^{++}_{cc}\rightarrow \Xi^+_c \pi^+$ \cite{Aaij:2018gfl}.
The weighted average of the $ \Xi^{++}_{cc} $ mass of the two decay modes was determined to be
$3621.24 \pm 0.65 (\text{stat.})\pm 0.31 (\text{syst.})~\text{MeV}/c^2 $~\cite{Aaij:2018gfl}.
The lifetime of the  $ \Xi^{++}_{cc} $ baryon was measured to be
$0.256\,^{+0.024}_{-0.022}(\text{stat.})\pm
0.014(\text{syst.})~\text{ps}$~\cite{Aaij:2018wzf}.
Recently,
with a data sample corresponding to an integrated luminosity  of 1.7~\ensuremath{\mbox{fb}^{-1}}, 
the $\Xi^{++}_{cc}\rightarrow D^+p K^- \pi^+$ decay has been searched for by the LHCb
collaboration, 
but no signal was found~\cite{Aaij:2019dsx}. Certainly, experiments will
continue to seek to solve the unexpected difference in parameters of these states and  will also search for other doubly heavy particles.

 As can be seen, the result of the LHCb collaboration for  the mass of the $ \Xi_{cc} $ is about $100~ \text{MeV}/c^2 $ 
higher than the value  reported by the  SELEX collaboration. The difference between these two results has motivated theoretical research to investigate the origin of  this difference. In Ref. \cite{Brodsky:2017ntu},
the authors have shown that the SELEX and
the LHCb results for the production of doubly charmed baryons can both be correct
if supersymmetric algebra is applied to hadron spectroscopy, together with the intrinsic heavy-quark QCD mechanism for the hadroproduction of heavy hadrons at large $ x_F $.

On the theoretical side, studies on  doubly heavy baryons  are needed to provide many inputs to experiments. Some aspects of doubly heavy baryons have been  discussed in Refs.  \cite{Bagan:1992za,Zhang:2008rt,Wang:2010hs,Aliev:2012ru,Aliev:2012iv,Azizi:2014jxa,Wang:2017mqp,Wang:2017azm,
Gutsche:2017hux,Li:2017pxa,Xiao:2017udy,Sharma:2017txj,Ma:2017nik,Hu:2017dzi,Shi:2017dto,Yao:2018ifh,Zhao:2018mrg,Wang:2018lhz,
Liu:2018euh,Xing:2018lre,Dhir:2018twm,Berezhnoy:2018bde,Jiang:2018oak,Zhang:2018llc,Gutsche:2018msz,Shi:2019fph,Hu:2019bqj,Brodsky:2011zs,Yan:2018zdt, Ozdem:2018uue}.  
The mechanisms of production and decay  of such systems  have  also been of interest to researchers for
many years \cite{Baranov:1995rc,Berezhnoy:1998aa,Gunter:2001qy,Ma:2003zk,Li:2007vy,Yang:2007ep,Zhang:2011hi,Jiang:2012jt,Martynenko:2014ola,Brown:2014ena,Trunin:2016uks,Huan-Yu:2017emk,Niu:2018ycb,Yao:2018zze}.
The production of doubly heavy baryons can be divided into two steps. The first step is
the perturbative production of a heavy quark pair in the hard interaction.
In the second step  this pair is transformed to the baryon within the soft hadronization process. The doubly heavy baryons can participate in many interactions and processes. The fusion of two $\Lambda_{c}$ to produce $\Xi_{cc}^{++} n$  results in an energy release of about $12 \text{MeV}$  and the fusion of two $\Lambda_{b}$ baryons to $\Xi_{bb}^{0}n$ released about $138 \text{MeV}$. This  suggests that an experimental setup may be designed to allow this huge released  energy to be used,  although the very short lifetimes of $cc$ and $bb$ baryons may prevent practical applications at the present time \cite{Karliner}.   

In this study, we
investigate  the strong coupling constants among the doubly heavy spin-1/2 baryons and light pseudoscalar mesons, $\pi$, $K$,  $ \eta $ and $ \eta^\prime $, which is the extension of our previous work \cite{Olamaei:2020bvw}. In Ref. \cite{Olamaei:2020bvw}, we investigated only the symmetric $  \Xi_{QQ^\prime }$ and calculated its coupling constant with   $\pi $ mesons. In the present study, we investigate the strong coupling constants of the $  \Xi_{QQ^\prime } $, $  \Xi^{\prime}_{QQ^\prime }$, $  \Omega_{QQ^\prime}$  and  $ \Omega^{\prime}_{QQ^\prime}$ doubly heavy baryons with all  the light pseudoscalar mesons, $\pi$, $K$,  $ \eta $ and $ \eta^\prime $  with different charges. Here $ Q $ and $ Q^\prime $
can both be $ b $ or $ c $ quarks. 
 We use the well established non-perturbative method of light cone QCD sum rules (LCSR) in the calculations. 
 In the framework of LCSR,  which has been developed based on the standard technique of  the SVZ  sum rule method \cite{Shifman:1978bx}, the nonperturbative dynamics of the quarks and gluons in the baryons are described by the light-cone distribution amplitudes (DAs). The LCSR approach uses
operator product expansion (OPE) near the lightcone $ x^2\approx 0 $  instead of the short distance  $ x\approx 0 $,
and the nonperturbative matrix elements are parameterized by the light cone DAs, which are classified according to their twists \cite{Balitsky:1989ry,Khodjamirian:1997ub,Braun:1997kw}.

The rest of the paper is organized as follows. In  the next section, we describe the formalism and obtain the sum rules for the strong coupling constants under study. In Section~\ref{NA}, the numerical analysis and results are
presented. Section~\ref{SC}  is reserved for summary and concluding notes.

\section{Strong coupling constants among doubly heavy baryons and pseudoscalar mesons }\label{LH}

Before going to the details of the  calculations for the strong coupling constants, we take a look at the ground state of the doubly heavy baryons in the quark model.  In the case of the doubly heavy baryons having two identical heavy quarks, i.e. $Q=Q^{\prime}$, the two heavy quarks form a diquark system with spin 1. After adding the light quark spin, the whole baryon may have spin 1/2 ($\Xi_{QQ}$ and $\Omega_{QQ}$) or 3/2 ($\Xi^*_{QQ}$ and $\Omega^*_{QQ}$).    Here the interpolating current should be symmetric with respect to the exchange of  heavy quarks. 
 In the case of different heavy quarks ($Q\neq Q^{\prime}$),  in addition to the above case, the diquark portion can also have spin zero where together with the light quark, the total spin of the whole baryon will be 1/2, which obviously leads to  anti-symmetric interpolating currents  with respect to the exchange of the two heavy quarks. They are usually denoted by the  primed baryons  $\Xi^{\prime}_{bc}$ and $\Omega^{\prime}_{bc}$.

The main inputs in the sum rule method are interpolating currents, which are written based on the general properties of the baryons and in terms of their quark contents.   In the case of doubly heavy baryons, the symmetric and anti-symmetric interpolating fields for spin-1/2 particles are given as:
\begin{eqnarray}\label{etaS}
\eta^{\cal S}&=&\frac{1}{\sqrt{2}}\epsilon_{abc}\Bigg\{(Q^{aT}Cq^b)\gamma_{5}Q'^c+
(Q'^{aT}Cq^b)\gamma_{5}Q^c+t (Q^{aT}C\gamma_{5}q^b)Q'^c \nonumber \\
&&+t(Q'^{aT}C
\gamma_{5}q^b)Q^c\Bigg\},
\end{eqnarray}
\begin{eqnarray}\label{etaA}
\eta^{\cal A}&=&\frac{1}{\sqrt{6}}\epsilon_{abc}\Bigg\{2(Q^{aT}CQ'^b)\gamma_{5}q^c+
(Q^{aT}Cq^b)\gamma_{5}Q'^c-(Q'^{aT}Cq^b)\gamma_{5}Q^c \nonumber\\
&&+2t (Q^{aT}C
\gamma_{5}Q'^b)q^c
+ t(Q^{aT}C\gamma_{5}q^b)Q'^c-t(Q'^{aT}C\gamma_{5}q^b)Q^c\Bigg\},
\end{eqnarray}
where $C$ stands for the charge conjugation operator, $T$ denotes the transposition and $t$ is an arbitrary mixing parameter where the case $t=-1$ corresponds to the Ioffe current. $Q^{(')}$ and $q$ stand for the heavy and light quarks respectively and $a$, $b$, and $c$ are the color indices. The quark contents for different members are shown in Table \ref{tab:baryon}.

 As an example, we demonstrate how the current of the doubly heavy baryons in its antisymmetric form is constructed considering all the quantum numbers. The simplest way of constructing a spin-$1/2$ baryon interpolating current is to make a diquark state of isospin and spin zero from two of three constituent quarks  with the third quark of isospin and spin-$1/2$. To make a diquark, we start with a meson interpolating current,
\begin{equation}
\eta_{meson}=\bar{q_1}\Gamma q_2~,
\end{equation}
where $ \Gamma={I,\gamma_5,\gamma_\mu,\gamma_5 \gamma_\mu,\sigma_{\mu \nu}} $. Then we replace the antiquark with its charge conjugation analog, where  $ q=C\bar{q}^T $. Therefore $ \bar{q}=q^TC $, in which $ C $ is the charge conjugation operator and $ C^T=C^{-1}=C^\dagger=-C $. This leads to the diquark interpolating currents 
\begin{equation}
\eta_{diquark}=q_1^T C\Gamma q_2~.
\end{equation}

Adding the third quark spinor to make the baryon current,  $ [q_1^T C\Gamma q_2] \Gamma' q_3 $, the generic form of the antisymmetric interpolating current for the doubly heavy baryons would be:
\begin{equation}\label{jj}
\eta^A\sim\epsilon_{abc}\big\{ (Q^{aT} C\Gamma Q'^b) \Gamma' q^c +(Q^{aT} C\Gamma q^b) \Gamma' Q'^c+(q^{aT} C\Gamma Q^b) \Gamma' Q'^c -\big( Q \leftrightarrow Q' \big)\big\},
\end{equation}

where $ \epsilon_{abc} $ makes the whole current color singlet.
To determine $ \Gamma$ and $\Gamma' $, we focus on the diquark part of the first term of the above equation. After transposing it we have:
\begin{equation}
[\epsilon_{abc} Q^{aT} C\Gamma Q'^b ]^T =-\epsilon_{abc} Q'^{bT} \Gamma^T C^{-1}  Q^a= \epsilon_{abc} Q'^{bT} C(C\Gamma^T C^{-1}) Q^a.
\end{equation}

Here we consider  $C^T = C^{-1}$, $C^2 = -1$ and the fact that the Grassmann numbers in the spinor components anticommute.
For the quantity $ C\Gamma^T C^{-1} $ we have:
\begin{equation}
C\Gamma^T C^{-1} =
\begin{cases}
\Gamma & \text{for $\Gamma = 1, \gamma_5, \gamma_{\mu}\gamma_5 $~,}\\
-\Gamma & \text{for $\Gamma = \gamma_{\mu}, \sigma_{\mu\nu}$~.}
\end{cases}       
\end{equation}

On the LHS, we switch the color dummy indices and get: 
\begin{equation}
[\epsilon_{abc} Q^{aT} C\Gamma Q'^b ]^T =\pm \epsilon_{abc} Q'^{aT} C\Gamma  Q^b,
\end{equation}

where the $+$ and $-$ signs are for $\Gamma = \gamma_{\mu}, \sigma_{\mu \nu}$ and $ \Gamma=1, \gamma_5,  \gamma_5\gamma_\mu$ respectively.
For the antisymmetric interpolating current, the RHS of the above equation is antisymmetric under $ Q\leftrightarrow Q' $ exchange and 
we have:
\begin{equation}\label{AntSymm}
[\epsilon_{abc} Q^{aT} C\Gamma Q'^b ]^T = \pm \epsilon_{abc} Q^{aT} C\Gamma  Q'^b~,
\end{equation}

where the $+$ and $-$ signs are for $ \Gamma=1, \gamma_5,  \gamma_5\gamma_\mu$ and  $\Gamma = \gamma_{\mu}, \sigma_{\mu \nu}$ respectively.
On the other hand, as $\epsilon_{abc} Q^{aT} C\Gamma Q'^b$ is a $1 \times 1$ matrix, it is equal to its transpose and therefore one can conclude that the only choices for $\Gamma$ matrices are $\Gamma = 1, \gamma_5, \gamma_5\gamma_\mu$.

As mentioned above, the simplest way of constructing spin-$1/2$ baryons is to suppose that the baryon spin  be equal to that of light quark $q$ (for the first term in Eq. (\ref{jj})) and thus the diquark part has a scalar structure, which implies that $\Gamma = 1, \gamma_5$.
Therefore, the allowed forms of the antisymmetric interpolating current may take just the following two forms: $\epsilon_{abc} (Q^{aT} C Q'^b)\Gamma' q^c $ and $ \epsilon_{abc} (Q^{aT} C \gamma_5 Q'^b)\Gamma' q^c $.

The matrices $\Gamma'_1$ and $\Gamma'_2$ can be determined considering Lorentz and parity symmetries. As the whole interpolating current is a Lorentz scalar,  there are two possibilities: $1$ and $\gamma_5$. The parity property of the interpolating current finally says that $\Gamma'_1 = \gamma_5$ and $\Gamma'_2 = 1$. 
Writing their linear combination as the most general form, the antisymmetric form of the  first term of Eq.~(\ref{jj}) is:
\begin{equation}\label{term1}
\eta^A\sim \epsilon_{abc}\big\{ (Q^{aT} C Q'^b) \gamma_5 q^c + t(Q^{aT} C \gamma_5 Q'^b)q^c -\big( Q \leftrightarrow Q' \big) \big\},
\end{equation}

where $  t$ is an arbitrary mixing parameter. Considering the Grassmann nature of the heavy quark spinor components, the antisymmetric property of $\epsilon_{abc}$ and $C^T = -C$, one can find out that the $-\big( Q \leftrightarrow Q' \big)$ terms are exactly the same as the first two, which yields:
\begin{eqnarray}
\eta^A\sim 2 \epsilon_{abc}\big\{ (Q^{aT} C Q'^b) \gamma_5 q^c + t(Q^{aT} C \gamma_5 Q'^b)  q^c \big\}.
\end{eqnarray}  
 
A similar argument can be used to calculate the second and third terms in Eq. (\ref{jj}). The symmetric interpolating current can be obtained in the same way but with the exception that in the exchange of heavy quarks in Eq. (\ref{AntSymm}) no minus sign is considered.

The main goal in this section is to find the strong coupling constants among the doubly heavy baryons with spin-1/2, 
$  \Xi_{QQ^\prime }$   $ \Xi^\prime_{QQ^\prime} $, $ \Omega_{QQ^\prime}$ and  $ \Omega^\prime_{QQ^\prime} $, with the light pseudoscalar mesons $ \pi $, $ K $,  $\eta$ and $ \eta^\prime $. To this end, we use the LCSR
approach as  one of the most powerful non-perturbative methods which is based on the light-cone OPE.
The starting point is to write the corresponding correlation function (CF):
\begin{eqnarray}\label{equ1} 
\Pi(p,q)= i \int d^4x e^{ipx} \left< {\cal P}(q) \vert {\cal T} \left\{
\eta (x) \bar{\eta} (0) \right\} \vert 0 \right>~,
\end{eqnarray}

where the two time-ordered interpolating currents of doubly-heavy baryons are sandwiched between the QCD vacuum and the on-shell pseudoscalar meson $ {\cal P}(q) $.  Here,  $p$ is the external four-momentum of the outgoing doubly heavy baryon. As the theory is translationally invariant we can choose one of the interpolating currents, $\bar{\eta}(0)$, at the origin.
It is worth noting again that in the symmetric interpolating current ($ \eta^{\cal S} $), heavy quarks may be identical or different whereas in the anti-symmetric one ($ \eta^{\cal A} $) they must be different.
\begin{table}
\setlength{\tabcolsep}{1.5em} 
\centering
\begin{tabular}{cccc}
\hline
\hline
\textcolor{red}{Baryon}    &\textcolor{red}{ $ q $ }& \textcolor{red}{$ Q $ }& \textcolor{red}{$ Q^\prime $ }\\
\hline
\hline
\\
$  \Xi_{QQ^\prime }$ or $ \Xi^\prime_{QQ^\prime} $     & $ u $ or $ d $    & $b  $   or $ c $ & $b  $   or $ c $    \\
          &         &  &     \\
$ \Omega_{QQ^\prime}$ or $ \Omega^\prime_{QQ^\prime} $      & $ s $   & $b  $   or $ c $ & $b  $   or $ c $       \\
    \\
\hline
\end{tabular}
\caption{Quark contents of the doubly heavy spin-1/2 baryons.} \label{tab:baryon} 
\end{table}

In the LCSR approach, the cornerstone is the CF. It can be calculated in two different ways. In the timelike region, one can insert the complete set of hadronic states with the same quantum numbers as the interpolating currents to extract and isolate the ground states. It is called the phenomenological or physical side of the CF. In the spacelike region which is free of singularities, one can calculate the CF in terms of QCD degrees of freedom using OPE. It is known as the QCD or theoretical side. These two representations, which respectively are the real and imaginary parts of the CF, can be matched via a dispersion relation to find the corresponding sum rule. The divergences coming from the dispersion integral, as well as higher states and continuum, are suppressed using the well-known method of Borel transformation and continuum subtraction.

On the phenomenological side, after inserting the  complete sets of
hadronic states with the same quantum numbers as the interpolating currents and performing the Fourier integration over $x$, we get
\begin{eqnarray}\label{phside} 
\Pi^{\text{Phys.}}(p,q)=\frac{\langle 0\vert \eta\vert B_2(p,r)\rangle  \langle B_2(p,r){\cal P}(q)\vert B_1(p+q,s)\rangle\langle B_1(p+q,s) \vert \bar{\eta}\vert 0\rangle}{(p^2-m_1^2)[(p+q)^2-m_2^2]} +\cdots~,
\end{eqnarray}

where the ground states are isolated and the dots represent the contribution of the  higher states and continuum. $B_1(p+q,s)$ and $B_2(p,r)$ are the initial and final doubly heavy baryons with spins $s$ and $r$ respectively.
The matrix element $\langle 0\vert \eta\vert B_i(p,s)\rangle$ is defined as: 
\begin{eqnarray}\label{me1} 
\langle 0\vert \eta\vert B_i(p,s)\rangle &=&\lambda_{B_i}u(p,s),
\end{eqnarray}

where $\lambda_{B_i}$ are the residues and $u(p,s)$ is the Dirac spinor for the baryons $B_i$ with momentum $p$ and spin $s$.
By the Lorentz and parity consideration,  one can write the matrix element $\langle B_2(p,r){\cal P}(q)\vert B_1(p+q,s)\rangle$ in terms of the strong coupling constant and Dirac spinors as:
\begin{eqnarray}\label{me2} 
\langle B_2(p,r){\cal P}(q)\vert B_1(p+q,s)\rangle &=& g_{B_1 B_2{\cal P}}
\bar{u}(p,r)\gamma_5 
u(p+q,s) ~,
\end{eqnarray}

where $ g_{B_1 B_2{\cal P}} $, represents the strong coupling constant for the strong decay $ B_1\rightarrow B_2 {\cal P}$. The final expression for the phenomenological side of the correlation function is obtained by inserting Eqs.~(\ref{me1}) and (\ref{me2}) into Eq.~(\ref{phside}) and summing over spins:
\begin{eqnarray}\label{CFPhys}
\Pi^{\text{Phys.}}(p,q)=\frac{ g_{B_1 B_2{\cal P}}\lambda_{B_1}\lambda_{B_2}}{(p^2-m_{B_2}^2)[(p+q)^2-m_{B_1}^2]}[\rlap/q \rlap/p\gamma_5 +\cdots~] +\cdots,
\end{eqnarray}

where the ellipsis inside the bracket denote several $\gamma$-matrix structures that may appear in the final expression due to the spin summation.  Here we select the structure $\rlap/q \rlap/p\gamma_5$ to perform analysis. 

To kill the higher states and continuum contributions we apply the double Borel transformation with respect to the square of the doubly heavy baryon momenta
$ p^2_1=(p+q)^2 $ and $ p^2_2=p^2 $, which leads to 
\begin{eqnarray}\label{CFPhysB}
{\cal B}_{p_1}(M_1^2){\cal B}_{p_2}(M_2^2)\Pi^{\text{Phys.}}(p,q) &\equiv & \Pi^{\text{Phys.}}(M^2)\nonumber\\
&=& g_{B_1 B_2 {\cal P}} \lambda_{B_1} \lambda_{B_2} e^{-m_{B_1}^2/M_1^2} e^{-m_{B_2}^2/M_2^2}\rlap/q \rlap/p\gamma_5~ + \cdots~,
\end{eqnarray}

where  $ M^2_1 $ and $ M^2_2 $ are the Borel parameters correspondinf to the square momenta $p_1^2$ and $p_2^2$ respectively, and $ M^2= M^2_1 M^2_2/(M^2_1+M^2_2) $. As the masses of the initial and final state baryons are the same or to a good approximation equal, the Borel parameters are chosen to be equal and therefore $ M^2_1 = M^2_2 = 2M^2$.

On the QCD side, choosing the corresponding structure to Eq. (\ref{CFPhys}) one can express the CF function as
\begin{equation}\label{QCD1}
\Pi^{\text{QCD}}(p,q)=\Pi\big(p,q\big)  \rlap/q \rlap/p\gamma_5,
\end{equation}

where $\Pi\big(p,q\big)$ is an invariant function of $(p+q)^2$ and $p^2$.  The main aim in this part is to determine this function in the Borel scheme. To this end,  we insert the explicit forms of the
interpolating currents (\ref{etaS}) and (\ref{etaA}) into the correlation function (\ref{equ1}) 
and use the Wick theorem to contract all the heavy quark fields. The result for the symmetric interpolating current is as follows:

\begin{eqnarray}\label{QCD2S}
\Pi^{\text{QCD}}_{(\cal{S})\rho\sigma}(p,q) &=& \frac{i}{2}\epsilon_{abc}\epsilon_{a'b'c'} \int d^4 x e^{i q.x} \langle {\cal P}(q) \vert \bar{q}^{c^\prime}_{\alpha}(0)q^{c}_{\beta}(x)\vert 0\rangle \Bigg\{ \Bigg[ \Big(\tilde{S}^{aa^{\prime}}_{Q}(x) \Big)_{\alpha\beta}\Big( \gamma_5 S^{bb^{\prime}}_{Q^{\prime}}(x) \gamma_5\Big)_{\rho\sigma}\nonumber \\
&+& \Big(\gamma_5 S_{Q^{\prime}}^{bb^{\prime}}(x)C\Big)_{\rho\alpha}\Big(CS^{aa^{\prime}}_{Q}(x)\gamma_5 \Big)_{\beta\sigma} + t\Big\{\Big(\gamma_5 \tilde{S}_{Q}^{aa^{\prime}}(x)\Big)_{\alpha\beta}\Big(\gamma_5 S^{bb^{\prime}}_{Q^{\prime}}(x) \Big)_{\rho\sigma} \nonumber\\
&+& \Big( \tilde{S}_{Q}^{aa^{\prime}}(x)\gamma_5\Big)_{\alpha\beta}\Big(S^{bb^{\prime}}_{Q^{\prime}}(x)\gamma_5 \Big)_{\rho\sigma} +\Big(\gamma_5 S_{Q^{\prime}}^{bb^{\prime}}(x)C\gamma_5\Big)_{\rho\alpha}\Big(CS^{aa^{\prime}}_{Q}(x)\Big)_{\beta\sigma} \nonumber\\
&-&\Big( S_{Q^{\prime}}^{bb^{\prime}}(x)C\Big)_{\rho\alpha}\Big(\gamma_5CS^{aa^{\prime}}_{Q}(x)\gamma_5\Big)_{\beta\sigma}
 \Big\}+ t^2 \Big\{\Big(\gamma_5 \tilde{S}_{Q}^{aa^{\prime}}(x)\gamma_5\Big)_{\alpha\beta}\Big(S^{bb^{\prime}}_{Q^{\prime}}(x)\Big)_{\rho\sigma} \nonumber\\
  &-& \Big( S_{Q^{\prime}}^{bb^{\prime}}(x)C\gamma_5\Big)_{\rho\alpha}\Big(\gamma_5CS^{aa^{\prime}}_{Q}(x)\Big)_{\beta\sigma} \Big\}\Bigg]+\Bigg(Q \longleftrightarrow Q^{\prime}\Bigg)\Bigg\},
\end{eqnarray}

where the $\rho$ and $\sigma$ are Dirac indices which run through 1 to 4, $S_{Q}^{aa^{\prime}}(x)$ is the heavy quark propagator, $\tilde{S}=C S^T C$ and the subscript ${\cal S}$ denotes the symmetric part. $\langle {\cal P}(q) \vert \bar{q}^{c^\prime}_{\alpha}(x)q^{c}_{\beta}(0)\vert 0\rangle$ are the non-local matrix elements for the light quark contents of the doubly heavy baryons and are purely non-perturbative.
For the anti-symmetric part we have:

\begin{eqnarray}\label{QCD2A}
\Pi^{\text{QCD}}_{(\cal A)\rho\sigma}(p,q) &=& \frac{i}{6}\epsilon_{abc}\epsilon_{a'b'c'} \int d^4 x e^{i q.x} \langle {\cal P}(q) \vert \bar{q}^{c^\prime}_{\alpha}(0)q^{c}_{\beta}(x)\vert 0\rangle \Bigg\{ 4 \text{Tr}\big[ \tilde{S}^{aa^{\prime}}_{Q}(x)S^{bb^{\prime}}_{Q^{\prime}}(x)\big]\gamma^{5}_{\alpha\sigma}\gamma^{5}_{\rho\beta}\nonumber\\
&-&2\Big( \tilde{S}^{aa^{\prime}}_{Q}(x)S^{bb^{\prime}}_{Q^{\prime}}(x)\gamma^5\Big)_{\alpha\sigma}\gamma^{5}_{\rho\beta} -2\Big(\gamma^5  S^{bb^{\prime}}_{Q^{\prime}}(x)\tilde{S}^{aa^{\prime}}_{Q}(x) \Big)_{\rho\beta}\gamma^5_{\alpha\sigma}\nonumber\\
&-&2\Big(\tilde{S}^{bb^{\prime}}_{Q^{\prime}}(x)S^{aa^{\prime}}_{Q}(x)\gamma^5\Big)_{\alpha\sigma}\gamma^5_{\rho\beta}-2\Big(\gamma^5S^{aa^{\prime}}_{Q}(x)\tilde{S}^{bb^{\prime}}_{Q^{\prime}}(x)\Big)_{\rho\beta}\gamma^5_{\alpha\sigma} \nonumber\\
 &+&\Big(\tilde{S}^{aa^{\prime}}_{Q}(x)\Big)_{\alpha\beta}\Big(\gamma^5S^{bb^{\prime}}_{Q^{\prime}}(x)\gamma^5 \Big)_{\rho\sigma} +\Big(\tilde{S}^{bb^{\prime}}_{Q^{\prime}}(x)\Big)_{\alpha\beta}\Big(\gamma^5S^{aa^{\prime}}_{Q}(x)\gamma^5\Big)_{\rho\sigma}\nonumber\\
&+& \Big(\gamma^5S^{bb^{\prime}}_{Q^{\prime}}(x)C\Big)_{\rho\alpha}\Big(CS^{aa^{\prime}}_{Q}(x)\gamma^5 \Big)_{\beta\sigma}+\Big(\gamma^5S^{aa^{\prime}}_{Q}(x)C\Big)_{\rho\alpha}\Big(CS^{bb^{\prime}}_{Q^{\prime}}(x)\gamma^5\Big)_{\beta\sigma} \nonumber\\
 &+&t\Bigg[ 4\text{Tr}\big[\tilde{S}^{aa^{\prime}}_{Q}(x)S^{bb^{\prime}}_{Q^{\prime}}(x)\gamma^5\big]\gamma^5_{\rho\beta}\delta_{\alpha\sigma} +4\text{Tr}\big[S^{bb^{\prime}}_{Q^{\prime}}(x)\tilde{S}^{aa^{\prime}}_{Q}(x)\gamma^5\big]\gamma^5_{\alpha\sigma}\delta_{\rho\beta}\nonumber\\
 &+&2\Big(\tilde{S}^{bb^{\prime}}_{Q^{\prime}}(x)\gamma^5S^{aa^{\prime}}_{Q}(x)C\gamma^5\Big)_{\alpha\sigma}\delta_{\beta\rho} -2\Big(S^{aa^{\prime}}_{Q}(x)\tilde{S}^{bb^{\prime}}_{Q^{\prime}}(x)\gamma^5\Big)_{\beta\rho}\gamma^5_{\alpha\sigma} \nonumber\\
 &-&2\Big(\gamma^5  \tilde{S}^{aa^{\prime}}_{Q}(x)S^{bb^{\prime}}_{Q^{\prime}}(x) \Big)_{\alpha\sigma}\gamma^5_{\rho\beta} -2\Big(\gamma^5S^{bb^{\prime}}_{Q^{\prime}}(x)\gamma^5 \tilde{S}^{aa^{\prime}}_{Q}(x)\Big)_{\rho\beta}\delta_{\sigma\alpha} \nonumber \\
&-&2\Big(\gamma^5\tilde{S}^{bb^{\prime}}_{Q^{\prime}}(x)S^{aa^{\prime}}_{Q}(x)\Big)_{\alpha\sigma}\gamma^5_{\rho\beta}-2\Big(\gamma^5S^{aa^{\prime}}_{Q}(x)\gamma^5\tilde{S}^{bb^{\prime}}_{Q^{\prime}}(x)\Big)_{\rho\beta}\delta_{\alpha\sigma}\nonumber\\
&-&2\Big(\tilde{S}^{aa^{\prime}}_{Q}(x)\gamma^5S^{bb^{\prime}}_{Q^{\prime}}(x)\gamma^5 \Big)_{\alpha\sigma}\delta_{\rho\beta} -2\Big(S^{bb^{\prime}}_{Q^{\prime}}(x)\tilde{S}^{aa^{\prime}}_{Q}(x)\gamma^5\Big)_{\rho\beta}\gamma^5_{\alpha\sigma}\nonumber\\
 &+&\Big(\gamma^5\tilde{S}^{aa^{\prime}}_{Q}(x)\Big)_{\alpha\beta}\Big(\gamma^5 S^{bb^{\prime}}_{Q^{\prime}}(x)\Big)_{\rho\sigma} +\Big(\gamma^5S^{bb^{\prime}}_{Q^{\prime}}(x)C\gamma^5\Big)_{\rho\alpha}\Big(C S^{aa^{\prime}}_{Q}(x)\Big)_{\beta\sigma}\nonumber\\
 &+&\Big(\tilde{S}^{aa^{\prime}}_{Q}(x)\gamma^5\Big)_{\alpha\beta}\Big(S^{bb^{\prime}}_{Q^{\prime}}(x)\gamma^5\Big)_{\rho\sigma} +\Big(\gamma^5CS^{aa^{\prime}}_{Q}(x)\gamma^5\Big)_{\beta\sigma}\Big(S^{bb^{\prime}}_{Q^{\prime}}(x)C\Big)_{\rho\alpha} \nonumber\\
 &+&\Big(S^{aa^{\prime}}_{Q}(x)C\Big)_{\rho\alpha}\Big(\gamma^5CS^{bb^{\prime}}_{Q^{\prime}}(x)\gamma^5\Big)_{\beta\sigma} +\Big(\tilde{S}^{bb^{\prime}}_{Q^{\prime}}(x)\gamma^5\Big)_{\alpha\beta}\Big(S^{aa^{\prime}}_{Q}(x)\gamma^5\Big)_{\rho\sigma}
  \nonumber \\
 &+&\Big(\gamma^5S^{aa^{\prime}}_{Q}(x)C\gamma^5\Big)_{\rho\alpha}\Big( CS^{bb^{\prime}}_{Q^{\prime}}(x) \Big)_{\beta\sigma} +\Big(\gamma^5\tilde{S}^{bb^{\prime}}_{Q^{\prime}}(x)\Big)_{\alpha\beta}\Big(\gamma^5S^{aa^{\prime}}_{Q}(x)\Big)_{\rho\sigma} \Bigg] \nonumber\\
 &+& t^2\Bigg[ 4\text{Tr}\big[\tilde{S}^{aa^{\prime}}_{Q}(x)\gamma^5S^{bb^{\prime}}_{Q^{\prime}}(x)\gamma^5\big] \delta_{\alpha\sigma}\delta_{\beta\rho} -2\Big(\gamma^5\tilde{S}^{aa^{\prime}}_{Q}(x)\gamma^5S^{bb^{\prime}}_{Q^{\prime}}(x)\Big)_{\alpha\sigma}\delta_{\rho\beta} \nonumber\\
 &+&2\Big(S^{aa^{\prime}}_{Q}(x)\gamma^5\tilde{S}^{bb^{\prime}}_{Q^{\prime}}(x)\gamma^5\Big)_{\rho\beta}\delta_{\alpha\sigma} -2\Big(S^{bb^{\prime}}_{Q^{\prime}}(x)\gamma^5\tilde{S}^{aa^{\prime}}_{Q}(x)\gamma^5\Big)_{\rho\beta}\delta_{\alpha\sigma} \nonumber \\ &+&2\Big(\gamma^5\tilde{S}^{bb^{\prime}}_{Q^{\prime}}(x)\gamma^5S^{aa^{\prime}}_{Q}(x)\Big)_{\alpha\sigma}\delta_{\beta\rho} +\Big(S^{bb^{\prime}}_{Q^{\prime}}(x)\Big)_{\rho\sigma}\Big(\gamma^5\tilde{S}^{aa^{\prime}}_{Q}(x)\gamma^5\Big)_{\alpha\beta} \nonumber \\ 
 &+& \Big(S^{aa^{\prime}}_{Q}(x)\Big)_{\rho\sigma}\Big(\gamma^5\tilde{S}^{bb^{\prime}}_{Q^{\prime}}(x)\gamma^5\Big)_{\alpha\beta} +\Big(S^{bb^{\prime}}_{Q^{\prime}}(x)C\gamma^5\Big)_{\sigma\alpha}\Big(\gamma^5CS^{aa^{\prime}}_{Q}(x)\Big)_{\beta\rho} \nonumber \\ &+&\Big(S^{aa^{\prime}}_{Q}(x)C\gamma^5\Big)_{\rho\alpha}\Big(\gamma^5CS^{bb^{\prime}}_{Q^{\prime}}(x)\Big)_{\beta\sigma}\Bigg]\Bigg\}.
\end{eqnarray}

There is also a symmetric-anti-symmetric form of the CF, $\Pi^{\text{QCD}}_{(\cal{S}\cal{A})}(p,q)$, which is the result of taking one interpolating current (say $\eta$) to be in the symmetric and the other ($\bar{\eta}$) in the anti-symmetric form, which represents the strong decays in which the initial baryon is anti-symmetric and the final one is symmetric. 

The explicit expression for the heavy quark propagator is  given as (see  Ref.~\cite{Balitsky:1987bk}):
\begin{eqnarray}\label{HQP}
S_Q^{aa^{\prime}}(x) &=& {m_Q^2 \over 4 \pi^2} {K_1(m_Q\sqrt{-x^2}) \over \sqrt{-x^2}}\delta^{aa^{\prime}} -
i {m_Q^2 \rlap/{x} \over 4 \pi^2 x^2} K_2(m_Q\sqrt{-x^2})\delta^{aa^{\prime}}\nnb \\& -&
ig_s \int {d^4k \over (2\pi)^4} e^{-ikx} \int_0^1
du \Bigg[ {\rlap/k+m_Q \over 2 (m_Q^2-k^2)^2} \sigma^{\mu\nu} G_{\mu\nu}^{aa^{\prime}} (ux)
 \nnb \\
&+&
{u \over m_Q^2-k^2} x^\mu  \gamma^\nu G_{\mu\nu}^{aa^{\prime}}(ux) \Bigg]+\cdots.
\end{eqnarray}

The first two terms are the free part or perturbative contributions where $K_1$ and $K_2$ are  modified Bessel functions of the second kind. Terms $\sim G_{\mu\nu}^{ab}$ are due to the expansion of the propagator on the light-cone and correspond to the interaction with the gluon field. Here we use the shorthand notation 
\begin{eqnarray}
G^{aa^{\prime}}_{\mu \nu }\equiv G^{A}_{\mu \nu }t^{aa^{\prime}}_{A},
\end{eqnarray}

with $A=1,\,2\,\ldots 8$ and $t_{A}=\lambda_{A}/2$  where $\lambda _{A}$ are the Gell-Mann matrices.

Inserting the heavy quark propagator (\ref{HQP}) into the CFs (\ref{QCD2S}) and (\ref{QCD2A}) would lead to several kinds of contributions each representing a different Feynman diagram. There are two heavy quark propagators in each term of the CFs.
The leading order contribution consists of a bare loop, depicted in Fig. \ref{fig:PertD}. To calculate that, every heavy quark propagator is replaced by its perturbative terms. The non-perturbative part of this contribution comes from the non-local matrix elements of the pseudoscalar meson which are defined in terms of distribution amplitudes (DAs) of twist two and higher.
\begin{figure}[h]
	\centering
	\includegraphics[width=0.47\textwidth]{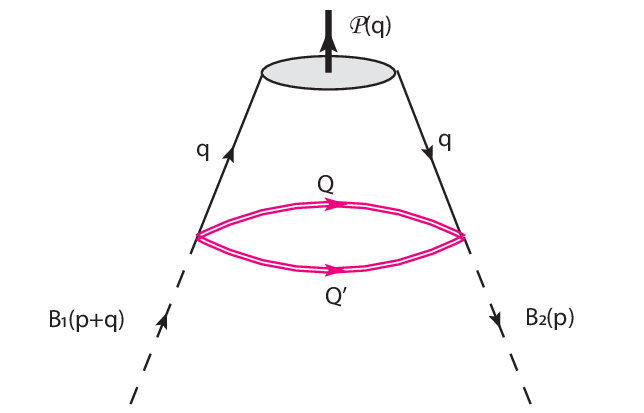}
	\caption{The leading order diagram contributing to $ \Pi(p,q) $.}
	\label{fig:PertD}
\end{figure}

Multiplication of the perturbative part of one heavy quark propagator and the gluon interaction part of another one leads to   contributions which can be calculated using the pseudoscalar meson three particle DAs. It is responsible for the exchange of one gluon between one of the heavy quarks and the outgoing meson, as shown in Fig. \ref{fig:Npert1}. 
\begin{figure}[h]
	\centering
	\includegraphics[width=0.47\textwidth]{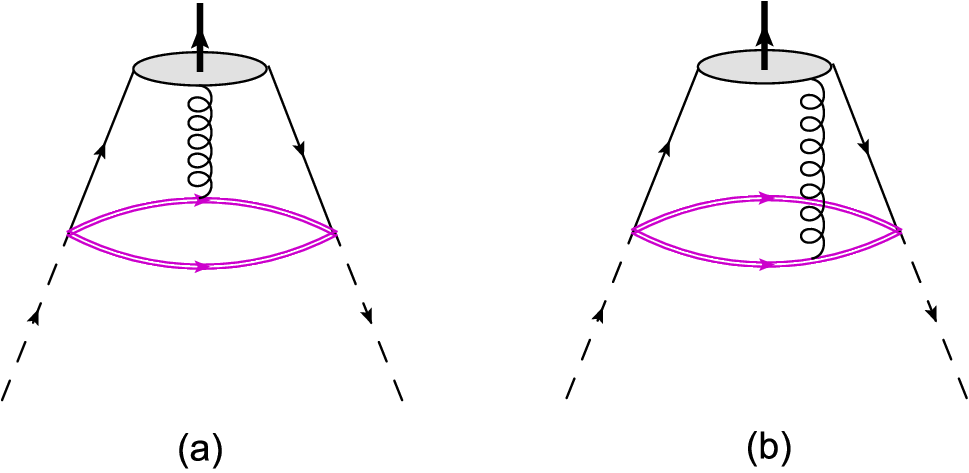}
	\caption{The one-gluon exchange diagrams.}
	\label{fig:Npert1}
\end{figure}
 The higher order contributions corresponding to at least four-particle DAs, which are not available yet,  are  neglected in this work. However,  we take into account the two-gluon condensates contributions $\sim g_s^2 \langle GG \rangle$. 
 To proceed we use the replacement
 \begin{equation}\label{sumcolor1}
 \overline{q}_{\alpha }^{c^{\prime }}(0)q_{\beta }^{c}(x)\rightarrow \frac{1}{%
 	3}\delta^{cc^{\prime }}\overline{q}_{\alpha }(0)q_{\beta }(x).
 \end{equation}
  which applies the projector onto the color singlet product of quark fields. One can decompose $\overline{q}_{\alpha }(x)q_{\beta }(0)$ into terms that have definite transformation properties under the Lorentz group and parity, using the completeness relation to get the expansion 
  \begin{equation}
  \overline{q}_{\alpha }(0)q_{\beta }(x)\equiv \frac{1}{4}\Gamma_{\beta \alpha }^{J}\overline{q}(0)\Gamma_{J}q(x),  \label{eq:Expan}
  \end{equation}

where $\Gamma^J$ runs over all possible $\gamma-$matrices with definite parity and Lorentz transformation property as
\begin{equation}\label{gammaexp}
\Gamma ^{J}=\mathbf{1,\ }\gamma _{5},\ \gamma _{\mu },\ i\gamma _{5}\gamma
_{\mu },\ \sigma _{\mu \nu }/\sqrt{2}.
\end{equation}

This helps us to project quarks onto the corresponding distribution amplitudes.

In the following, we would like to briefly explain how the contributions of, for instance ,the diagrams in Fig. \ref{fig:PertD} and Fig. \ref{fig:Npert1} are calculated.
For a symmetric current and Fig. \ref{fig:PertD}, we get
\begin{eqnarray}\label{QCD2Spert}
\Pi^{\text{QCD}(\text{1})}_{(\cal{S})\rho\sigma}(p,q) &=&   \frac{i}{4}\int d^4 x e^{i q.x} \langle {\cal P}(q) \vert \bar{q}(0)\Gamma^Jq(x)\vert 0\rangle \Bigg\{\Bigg[ \text{Tr}\big[ \Gamma_J\tilde{S}^{(\text{pert.})}_{Q}(x) \big]\Big( \gamma_5 S^{(\text{pert.})}_{Q^{\prime}}(x) \gamma_5\Big)_{\rho\sigma}\nonumber \\
&+& \Big(\gamma_5 S_{Q^{\prime}}^{(\text{pert.})}(x)\tilde{\Gamma}_JS_{Q}^{(\text{pert.})}(x)\gamma_5 \Big)_{\rho\sigma} + t\Big\{ \text{Tr}\big[\Gamma_J\gamma_5 \tilde{S}_{Q}^{(\text{pert.})}(x)\big]\Big(\gamma_5 S_{Q^{\prime}}^{(\text{pert.})}(x) \Big)_{\rho\sigma} \nonumber\\
&+& \text{Tr}\big[\Gamma_J \tilde{S}_{Q}^{(\text{pert.})}(x)\gamma_5\big]\Big(S_{Q^{\prime}}^{(\text{pert.})}(x)\gamma_5 \Big)_{\rho\sigma} +\Big(\gamma_5 S_{Q^{\prime}}^{(\text{pert.})}(x)\gamma_5\tilde{\Gamma}_JS_{Q}^{(\text{pert.})}(x)\Big)_{\rho\sigma} \nonumber\\
&-&\Big( S_{Q^{\prime}}^{(\text{pert.})}(x)\tilde{\Gamma}_J\gamma_5S_{Q}^{(\text{pert.})}(x)\gamma_5\Big)_{\rho\sigma}
\Big\} + t^2 \Big\{ \text{Tr}\big[\Gamma_J\gamma_5 \tilde{S}_{Q}^{(\text{pert.})}(x)\gamma_5\big]\Big(S_{Q^{\prime}}^{(\text{pert.})}(x)\Big)_{\rho\sigma} \nonumber\\
&-& \Big( S_{Q^{\prime}}^{(\text{pert.})}(x)\gamma_5\tilde{\Gamma}_J\gamma_5S_{Q}^{(\text{pert.})}(x)\Big)_{\rho\sigma} \Big\}\Bigg] + \Bigg( Q \leftrightarrow Q^{\prime}\Bigg) \Biggr\},
\end{eqnarray}

where $\tilde{\Gamma}_J=C\Gamma^{\text{T}}_JC$ and
\begin{eqnarray}\label{HQPpert}
S_Q^{(\text{pert.})}(x) &=& {m_Q^2 \over 4 \pi^2} {K_1(m_Q\sqrt{-x^2}) \over \sqrt{-x^2}} -
i {m_Q^2 \rlap/{x} \over 4 \pi^2 x^2} K_2(m_Q\sqrt{-x^2}).
\end{eqnarray}

 Fig. \ref{fig:Npert1} denotes contributions to  the CF due to the exchange of one gluon between the heavy quark $Q$ or $Q^{\prime}$ and the pseudoscalar meson $\cal{P}$. To be precise, taking the emission of the gluon from $Q$, Fig. \ref{fig:Npert1}a, one can write the corresponding CF by replacing $S^{bb^{\prime}}_{Q^{\prime}}(x)$ with its perturbative free part (\ref{HQPpert}) and $S^{aa^{\prime}}_{Q}(x)$ with its non-perturbative gluonic part
\begin{eqnarray}\label{HQPnp}
S^{aa^{\prime}(non-p.)}_{Q}(x)&=&-
ig_s \int {d^4k \over (2\pi)^4} e^{-ikx} \int_0^1
du G^{aa^{\prime}}_{\mu\nu}(ux) \Delta^{\mu\nu}_{Q}(x),
\end{eqnarray}
where $ \Delta^{\mu\nu}_{Q}(x)$ is defined to be:
\begin{eqnarray}\label{HQPgamma}
\Delta^{\mu\nu}_{Q}(x)&=& \dfrac{1}{2 (m_Q^2-k^2)^2}\Big[(\rlap/k+m_Q)\sigma^{\mu\nu} + 2u (m_Q^2-k^2)x^\mu \gamma^\nu\Big].
\end{eqnarray}

After summing over the color indices one can find the following relation for the symmetric current and   contribution of  the exchange of the gluon between the heavy quark $Q$ and the light pseudoscalar meson ${\cal P}$  (Fig. \ref{fig:Npert1}a):
\begin{eqnarray}\label{QCD2Snp}
\Pi^{\text{QCD(2a)}}_{({\cal S})\rho\sigma}(p,q) &=&- \frac{g_s}{12}\int d^4 x \int {d^4k \over (2\pi)^4}  \int_0^1
du  e^{i (q-k).x} \langle {\cal P}(q) \vert \bar{q}(x)\Gamma^J G_{\mu\nu}(ux)q(0)\vert 0\rangle \nonumber \\
&\times& \Bigg\{ \Bigg[ \text{Tr}\big[ \Gamma_J\tilde{\Delta}^{\mu\nu}_{Q}(x) \big]\Big( \gamma_5 S^{(\text{pert.})}_{Q^{\prime}}(x) \gamma_5\Big)_{\rho\sigma}+ \Big(\gamma_5 S_{Q^{\prime}}^{(\text{pert.})}(x)\tilde{\Gamma}_J\Delta^{\mu\nu}_{Q}(x)\gamma_5 \Big)_{\rho\sigma} \nonumber \\ &+& t\Big\{ \text{Tr}\big[\Gamma_J\gamma_5 \tilde{\Delta}^{\mu\nu}_{Q}(x)\big]\Big(\gamma_5 S_{Q^{\prime}}^{(\text{pert.})}(x) \Big)_{\rho\sigma} + \text{Tr}\big[\Gamma_J \tilde{\Delta}^{\mu\nu}_{Q}(x)\gamma_5\big]\Big(S_{Q^{\prime}}^{(\text{pert.})}(x)\gamma_5 \Big)_{\rho\sigma} \nonumber \\ 
&+& \Big(\gamma_5 S_{Q^{\prime}}^{(\text{pert.})}(x)\gamma_5\tilde{\Gamma}_J\Delta^{\mu\nu}_{Q}(x)\Big)_{\rho\sigma} - \Big( S_{Q^{\prime}}^{(\text{pert.})}(x)\tilde{\Gamma}_J\gamma_5\Delta^{\mu\nu}_{Q}(x)\gamma_5\Big)_{\rho\sigma}
\Big\} \nonumber\\ 
&+& t^2 \Big\{ \text{Tr}\big[\Gamma_J\gamma_5 \tilde{\Delta}^{\mu\nu}_{Q}(x)\gamma_5\big]\Big(S_{Q^{\prime}}^{(\text{pert.})}(x)\Big)_{\rho\sigma} - \Big( S_{Q^{\prime}}^{(\text{pert.})}(x)\gamma_5\tilde{\Gamma}_J\gamma_5\tilde{\Delta}^{\mu\nu}_{Q}(x)\Big)_{\rho\sigma} \Big\}\Bigg] \nonumber\\ 
&+& \Bigg( \Delta^{\mu\nu}_{Q}(x) \leftrightarrow S^{(\text{pert.})}_{Q^{\prime}}(x) \Bigg) \Bigg\}.
\end{eqnarray}

where $ \tilde{\Delta}^{\mu\nu}_{Q}(x) =C \Delta^{T,\mu\nu}_{Q}(x) C$.
The other contribution, $\Pi^{\text{QCD(2b)}}_{({\cal S})}$, which is responsible for the exchange of one gluon between the heavy quark $Q^{\prime}$ and the pseudoscalar meson $\cal P$, can simply be calculated by taking into account the perturbative part of  the $Q$-propagator and the one-gluon emission part of   the $Q^{\prime}$-propagator. Other contributions as well as  the anti-symmetric CF, are calculated in a similar way. The results are too lengthy to  present here.

From Eqs. (\ref{QCD2Spert}) and (\ref{QCD2Snp}) it is clear that the non-perturbative nature of the interaction is represented by the non-local matrix elements: 
\begin{eqnarray}\label{matel}
&\langle {\cal P}(q) \vert \bar{q}(x)\Gamma^J q(0)\vert 0\rangle~,&  \nonumber\\
&\langle {\cal P}(q) \vert \bar{q}(x)\Gamma^J G_{\mu\nu}(ux)q(0)\vert 0\rangle~,&
\end{eqnarray}

which can be expressed in terms of two- and three-particle DAs of different twists for the light pseudoscalar meson ${\cal P}$. The expressions for the above matrix elements in terms of DAs and the explicit form of the DAs are given in the Appendices (\ref{APA}) and (\ref{DAs}), respectively. 

Inserting the heavy quark propagators (\ref{HQP}) and the expressions for non-local matrix elements (\ref{matel}) from Appendix (\ref{APA}) into the CFs (\ref{QCD2Spert}) and (\ref{QCD2Snp}) one obtains the version of CF that is ready for performing the Fourier and Borel transformations as well as continuum subtraction. At this stage the CF contains several kinds of configurations with the general form:
\begin{eqnarray}\label{STR1}
T_{[~~,\alpha,\alpha\beta]}(p,q)&=& i \int d^4 x \int_{0}^{1} dv  \int {\cal D}\alpha e^{ip.x} \big(x^2 \big)^n  [e^{i (\alpha_{\bar q} + v \alpha _g) q.x} \mathcal{G}(\alpha_{i}) , e^{iq.x} f(u)] \nonumber\\
&\times& [1 , x_{\alpha} , x_{\alpha}x_{\beta}]  K_{\mu}(m_Q\sqrt{-x^2})  K_{\nu}(m_Q\sqrt{-x^2}).
\end{eqnarray} 

Here the expressions in the brackets represent different structures that come from calculations. The blank subscript bracket indicates no $x_\alpha$ in the structure and $n$ is an integer. The two and three-particle matrix elements lead to wave functions that are denoted by $f(u)$ and $\mathcal{G}(\alpha_{i})$  respectively.
${\cal D}\alpha$ is called the measure and is defined:
\begin{equation*}
\int \mathcal{D}\alpha =\int_{0}^{1}d\alpha _{q}\int_{0}^{1}d\alpha _{\bar{q}%
}\int_{0}^{1}d\alpha _{g}\delta (1-\alpha _{q}-\alpha _{\bar{q}}-\alpha
_{g}).
\end{equation*}

There are several representations for the Bessel function $K_\nu$. Here we use the cosine representation as 
\begin{equation}\label{CosineRep}
K_\nu(m_Q\sqrt{-x^2})=\frac{\Gamma(\nu+ 1/2)~2^\nu}{\sqrt{\pi}m_Q^\nu}\int_0^\infty dt~\cos(m_Qt)\frac{(\sqrt{-x^2})^\nu}{(t^2-x^2)^{\nu+1/2}},
\end{equation}
which helps us to increase the radius of convergence of the CF \cite{Azizi:2018duk}.
To perform the Fourier transformation we use the exponential representations of the $x-$structures:
\begin{eqnarray}\label{trick1}
(x^2)^n &=& (-1)^n \frac{d^n}{d \beta^n}\big(e^{- \beta x^2}\big)\arrowvert_{\beta = 0}, \nnb \\
x_{\alpha} e^{i P.x} &=& (-i) \frac{d}{d P^{\alpha}} e^{i P.x}.
\end{eqnarray} 
To be specific, one specific configuration that appears has the generic form 
\begin{eqnarray}\label{Z1}
{\cal Z}_{\alpha\beta}(p,q) &=& i \int d^4 x \int_{0}^{1} dv  \int {\cal D}\alpha e^{i[p+ (\alpha_{\bar q} + v \alpha _g)q].x} \mathcal{G}(\alpha_{i}) \big(x^2 \big)^n  \nonumber\\
&\times& x_\alpha x_\beta  K_{\mu}(m_Q\sqrt{-x^2})  K_{\nu}(m_Q\sqrt{-x^2}).
\end{eqnarray}

Using some variable changes and performing the double Borel transformation by employing 
\begin{equation} \label{Borel1}
{\cal B}_{p_1}(M_{1}^{2}){\cal B}_{p_2}(M_{2}^{2})e^{b (p + u q)^2}=M^2 \delta(b+\frac{1}{M^2})\delta(u_0 - u) e^{\frac{-q^2}{M_{1}^{2}+M_{2}^{2}}},
\end{equation}

in which  $u_0 = M_{1}^{2}/(M_{1}^{2}+M_{2}^{2})$ one can find the final Borel transformed result for the corresponding structure:
\begin{eqnarray}\label{STR4}
{\cal Z}_{\alpha\beta}(M^2) &=& \frac{i  \pi^2 2^{4-\mu-\nu} e^{\frac{-q^2}{M_1^2+M_2^2}}}{M^2 m_{Q_1}^{2\mu} m_{Q_2}^{2\nu}}\int  \mathcal{D}\alpha  \int_{0}^{1} dv \int_{0}^{1} dz \frac{\partial^n }{\partial \beta^n} e^{-\frac{m_1^2 \bar{z} + m_2^2 z}{z \bar{z}(M^2 - 4\beta)}} z^{\mu-1}\bar{z}^{\nu-1} (M^2 - 4\beta)^{\mu+\nu-1} \nonumber\\
&\times & \delta[u_0 - (\alpha_{q} + v \alpha_{g})]  \Big[ p_\alpha p_\beta + (v \alpha_{g} +\alpha_{q})(p_\alpha q_\beta +q_\alpha p_\beta ) + (v \alpha_{g} +\alpha_{q})^2 q_\alpha q_\beta \nonumber \\ 
&&+ \frac{M^2}{2}g_{\alpha\beta} \Big].
\end{eqnarray}

The details of the calculation can be found in Ref. \cite{Olamaei:2020bvw}.

According to Eq. (\ref{CFPhys}), we choose the structure $\rlap/q \rlap/p \gamma_{5}$ on the  QCD side as well. Therefore, one can write 
\begin{eqnarray}
\Pi^{\text{QCD}}_{{ B}_1 { B}_2 {\cal P}}(M^2) = \Pi_{{ B}_1 { B}_2 {\cal P}}(M^2) \rlap/q \rlap/p \gamma_{5}, 
\end{eqnarray}

where ${ B}_1$, ${B}_2$ and ${\cal P}$ represent the initial baryon,  final baryon and  pseudoscalar meson, respectively.
 The coefficient of the structure $\rlap/q \rlap/p \gamma_{5}$, i.e. $\Pi_{{ B}_1 {B}_2 {\cal P}}(M^2)$, is obtained considering all the contributions discussed above.
As an example,  we present the expression for the invariant function for the specific channel $\Omega_{bb}\rightarrow\Xi_{bb}\bar{K}^0$ after the Borel transformation, which reads:
\begin{eqnarray}
\Pi_{\Omega_{bb}\Xi_{bb}\bar{K}^0}(M^2) &=& \dfrac{e^{-\frac{m_{\bar{K}^0}^2}{4M^2}}}{6912\pi^2M^6m_b}\int_{0}^{1}dz\dfrac{e^{-\frac{m_b^2}{M^2 z \bar{z}}}}{z^2 \bar{z}^2}\nonumber \\ 
&\times&\Bigg\{ 72 m_b M^6 z \bar{z}^2 \Bigg( 3f_{\bar{K}^0}m_{\bar{K}^0}^2m_b(t^2-1)\big(m_b^2+2M^2z\bar{z}\big){\mathbb A}(u_0) \nonumber \\
&+&2M^2z\Big[ -6f_{\bar{K}^0}m_bM^2(t^2-1)\bar{z}\varphi_{\bar{K}^0}(u_0) \nonumber \\
&+&\mu_{\bar{K}^0}(\tilde \mu_{\bar{K}^0}^2 -1)\Big( 2m_b^2(1+t^2)+3M^2(t-1)^2 z\bar{z}\Big)\varphi_{\sigma}(u_0)\Big] \Bigg)\nonumber\\
&+&432m_b M^8 z\bar{z}^2\int_{0}^{1}dv\int \mathcal{D}\alpha \Bigg( f_{\bar{K}^0}m_{\bar{K}^0}^2m_b(t^2-1)\delta[u_0-(\alpha_{q}+v\alpha_{g})] \nonumber\\
&\times& \Big( (2v-1)z{\cal A}_\parallel (\alpha_i)+(2z-3){\cal V}_\parallel(\alpha_i) +2\bar{z}{\cal V}_\perp(\alpha_i)\Big) \\
&-&\mu_{\bar{K}^0}M^2(t-1)^2z\bar{z}\delta^{\prime}[u_0-(\alpha_{q}+v\alpha_{g})] {\cal T}(\alpha_i)\Bigg) \nonumber\\
&+& g_s^2\langle GG \rangle\Bigg[ -3f_{\bar{K}^0}m_{\bar{K}^0}^2(t^2-1)\Big( 2m_b^6-3m_b^4 M^2\bar{z}^2 -6m_b^2M^4z \bar{z}^3-6M^6z^2\bar{z}^4{\mathbb A}(u_0) \Big)\nonumber\\
&+&4M^2\bar{z}\Big[-3f_{\bar{K}^0}(t^2-1)z\Big(-2m_b^4-m_b^2M^2(5z-3)\bar{z}+6M^4z\bar{z}^3 \Big)\varphi_{\bar{K}^0}(u_0)\nonumber\\
&-&\mu_{\bar{K}^0}(\tilde \mu_{\bar{K}^0}^2 -1)m_b\Big( 2m_b^4(1+t^2)+m_b^2M^2\big[(1+t^2)(1-4z)-6t \big]z \nonumber\\
&+&M^4\big[ (1+t^2)(1-4z)-6t\big]z^2\bar{z}\varphi_{\sigma}(u_0)\Big) \Big]\nonumber\\
&+&6\int_{0}^{1}dv\int\mathcal{D}\alpha \Bigg(-f_{\bar{K}^0}m_{\bar{K}^0}^2M^2(t^2-1)\bar{z}\delta[u_0-(\alpha_{q}+v\alpha_{g})]\nonumber\\
&\times&\Big\{ (2v-1)\Big[m_b^4-m_b^2M^2z(1+2z)-M^4z^2(1+2z)\bar{z}\Big]{\cal A}_\parallel (\alpha_i) \nonumber\\
&+&\Big[ -4m_b^2+m_b^2M^2z(1+2z)+M^4z^2(1+z-2z^2) \Big]{\cal V}_\parallel(\alpha_i)+2m_b^4{\cal V}_\perp(\alpha_i) \Big\}\nonumber\\
&+&m_b M^4\mu_{\bar{K}^0}(t-1)2(2v-1)z\bar{z}(m_b^2+M^2z\bar{z})\delta^{\prime}[u_0-(\alpha_{q}+v\alpha_{g})] {\cal T}(\alpha_i) \Bigg) \Bigg]\Bigg\}.\nonumber
\end{eqnarray}

The invariant functions for other channels on the  QCD side are calculated in the same manner but  are  not presented here because of their very lengthy expressions. 

The next step is the  continuum subtraction. To this end,  we set the argument in  $ e^{-\frac{m_1^2 \bar{z} + m_2^2 z}{M^2 z \bar{z}}} $ (in the case of different heavy quarks) or  $ e^{-\frac{m_1^2}{M^2 z \bar{z}}} $ (in the case of identical heavy quarks) equal to  $s_0$, with  $s_0$ being the continuum threshold for higher states and continuum.  Here, $ m_1 $ and $ m_2 $ stand for the heavy quark masses. As a result,   the limits of $z$ are changed:
\begin{eqnarray}\label{zsubtraction}
\int_{0}^{1}dz \rightarrow \int_{z_{\text{min}}}^{z_{\text{max}}}dz,
\end{eqnarray}	

where for two different heavy quarks we get:
\begin{eqnarray}\label{zlimits}
z_{\text{max}(\text{min})}=\frac{1}{2s_0}\Big[(s_0+m_1^2-m_2^2)+(-)\sqrt{(s0+m_1^2-m_2^2)^2-4m_1^2s_0}\Big].
\end{eqnarray}

In the case of the heavy quarks, being the same we just need to  put $ m_1=m_2 $ in this result.

After performing continuum subtraction, the invariant function  becomes $s_0$-dependent. We also have  another auxiliary parameter $ t $ in the currents. Hence, in terms of the three auxiliary parameters we analyze the results numerically with respect to their variations, and after equating the coefficients of the selected structures from both the physical and QCD sides, we get the strong coupling constants as:
\begin{eqnarray}\label{SR}
g_{B_1 B_2  {\cal P}}(M^2,s_0,t)=\frac{1}{\lambda_{B_1}\lambda_{B_2}} e^{\frac{m_{B_1}^2+m_{B_2}^2}{2M^2}}\Pi_{B_1 B_2  {\cal P}}(M^2,s_0,t).
\end{eqnarray}

We analyze these sum rules numerically in the next section.

\section{Numerical results}\label{NA}
There are two sets of input parameters needed to perform the numerical analysis. One corresponds to the mass and decay constants of the  light pseudoscalar meson as well as the non-perturbative parameters coming from the light-cone DAs of different twists calculated at the renormalization scale $\mu=1\text{GeV}$. These parameters are collected in Tables \ref{tabmeson} and \ref{tabDAs}, respectively. The other set of input parameterscorresponds to the doubly heavy baryons masses and residues, which are  taken from Ref. \cite{Aliev:2012ru} and are presented in Table \ref{tabBaryon}.
\begin{table}[t]
	\renewcommand{\arraystretch}{1.3}
	\addtolength{\arraycolsep}{1pt}
	$$
	\rowcolors{2}{cyan!10}{white}
	\begin{tabular}{|c|c|c|c|}
	\rowcolor{cyan!30}
	\hline \hline
	\mbox{Parameters}         &  \mbox{Values  $[\text{MeV}]$ }  
	\\
	\hline\hline
	$ m_{c} $    &  $ 1.275^{+0.025}_{-0.035}~\mbox{GeV} $      \\
	$ m_b $    & $  4.18^{+0.04}_{-0.03}~\mbox{GeV} $      \\
	$ m_{\eta } $    &   $547.862\pm0.018 $  \\
	$ m_{\eta^{\prime} } $    &   $957.78\pm0.06 $  \\
	$ m_{K^0}  $  &   $497.648\pm0.022$    \\
	$ m_{K^{\pm}}  $  &   $493.677\pm0.013$    \\
	$  f_\pi $    & $  131  $     \\
	$  f_\eta $    & $  130  $     \\
	$  f_{\eta^{\prime}} $    & $  136  $     \\
	$  f_{K} $    & $  160  $     \\
	\hline \hline
	\end{tabular}
	$$
	\caption{Meson masses and leptonic decay constants along with  heavy quark masses \cite{Tanabashi:2018oca,Ball:2005vx,Ball:2004ye,Ball:1998je}.} \label{tabmeson} 
	\renewcommand{\arraystretch}{1}
	\addtolength{\arraycolsep}{-1.0pt}
\end{table} 

\begin{table}[t]
	\renewcommand{\arraystretch}{1.3}
	\addtolength{\arraycolsep}{1pt}
	$$
			\rowcolors{2}{cyan!10}{white}
	\begin{tabular}{|c|c|c|c|c|c|}
				\rowcolor{cyan!30}
	\hline \hline
	\text{meson}  &  $a_2$ & $\eta_3$ & $w_3$ & $\eta_4$ & $w_4$  
	\\
	\hline\hline
	 $\pi$  & 0.44 & $0.015$ & -3 & 10 & 0.2     \\

	$ $K$ $ & 0.16  & $0.015$ & -3 & 0.6 & 0.2     \\
	$\eta $ & 0.2 & 0.013 & -3 & 0.5 & 0.2     \\
	\hline \hline
	\end{tabular}
	$$
	\caption{Input parameters for twist 2, 3 and 4 DAs at the renormalization scale $\mu=1\text{GeV}$ \cite{Ball:2005vx,Ball:2004ye}.}  \label{tabDAs} 
	\renewcommand{\arraystretch}{1}
	\addtolength{\arraycolsep}{-1.0pt}
\end{table} 

\begin{table}[t]
	\renewcommand{\arraystretch}{1.3}
	\addtolength{\arraycolsep}{1pt}
	$$
	\rowcolors{2}{cyan!10}{white}
	\begin{tabular}{|c|c|c|c|}
	\rowcolor{cyan!30}
	\hline \hline
	\mbox{Baryon}         &  \mbox{Mass} $[\text{GeV}]$  & \mbox{Residue $[\text{GeV}^3]$}
	\\
	\hline\hline
	
	$ \Xi_{cc}  $    &  $ 3621.4\pm0.8~\mbox{MeV} $ \cite{Tanabashi:2018oca}  & $0.16\pm0.03$ \\
	$ \Xi_{bc}  $    &  $ 6.72\pm0.20 $ & $0.28\pm0.05$  \\
	$ \Xi^\prime_{bc}  $    &  $ 6.79\pm0.20$ & $0.3\pm0.05$  \\
	$ \Xi_{bb} $    &  $ 9.96\pm0.90  $  & $0.44\pm0.08$  \\
	$ \Omega_{cc}  $    &  $ 3.73\pm0.20 $  & $0.18\pm0.04$ \\
	$ \Omega_{bc}  $    &  $ 6.75\pm0.30 $  & $0.29\pm0.05$ \\
	$ \Omega^\prime_{bc}  $    &  $ 6.80\pm0.30 $ & $0.31\pm0.06$  \\
	$ \Omega_{bb} $    &  $ 9.97\pm0.90  $  & $0.45\pm0.08$ \\
	
	\hline \hline
\end{tabular}
$$
\caption{Baryon masses and residues   \cite{Aliev:2012ru}.}  \label{tabBaryon} 
\renewcommand{\arraystretch}{1}
\addtolength{\arraycolsep}{-1.0pt}
\end{table}

The sum rules for the strong coupling constants also depend  on three auxiliary parameters $M^2$,  $s_0$ and $t$  that should be fixed. To find the working intervals of these parameters, the standard prescriptions of the method are used: the variations of the results with respect to the changes in these parameters should be minimal. 
First of all we would like to fix the working region of $t$. Considering $t=\tan\theta$,  we plot the strong coupling constants with respect to $\cos\theta$. We choose $\cos\theta$ and vary it in the interval $ [-1,1] $ in order to explore $ t $  in all regions.  Our analysis shows that the variations of the  results are minimal in the intervals $0.5 \leq \cos\theta \leq 0.7$ and $-0.7 \leq \cos\theta \leq -0.5$   for all the strong decays under study.


The continuum threshold depends on the energy of the first excited state in each channel. Unfortunately, we have no experimental information on the first excited doubly heavy baryons. We choose it in the interval  $(m_B+0.3)^2\leq s_0\leq(m_B+0.7)^2 \text{GeV}^2$, where dependence of the results on $s_0$ is  weak. As examples, we show the dependence of $g_{\Omega^{(\prime)}_{QQ^{(\prime)}} \Omega^{(\prime)}_{QQ^{(\prime)}} \eta (\eta^\prime)}  $  on continuum threshold  in Fig~\ref{fig:s0}. From this figure, we see that the strong coupling constants   depend very weakly on the variation of the $s_0$  in the selected intervals.
\begin{figure}[h!]
\includegraphics[width=1.0\textwidth]{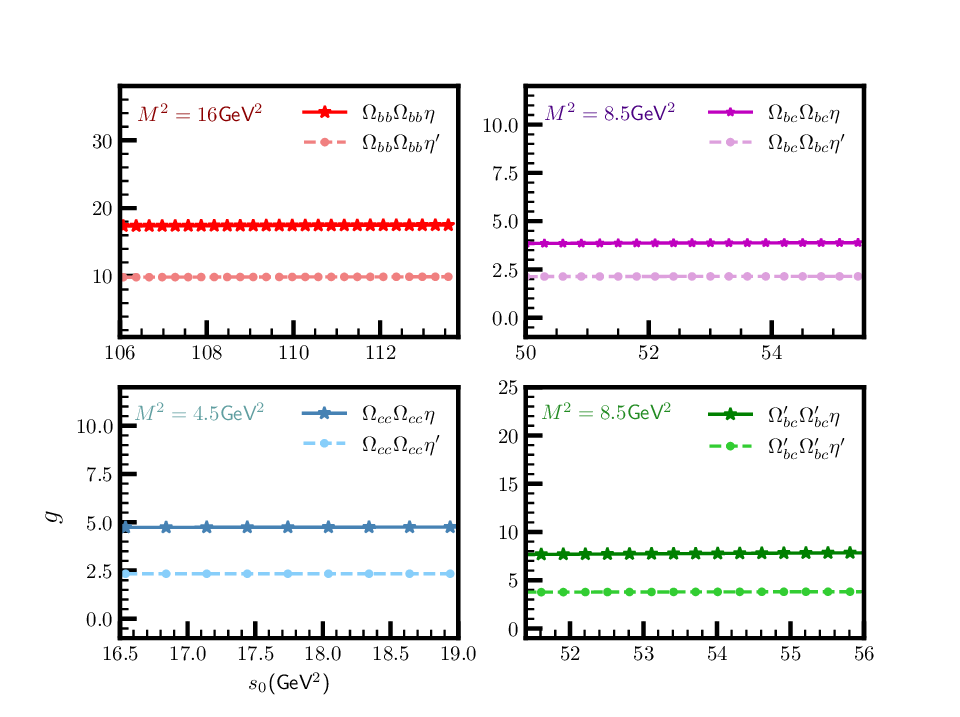}
\caption{The strong couplings $g_{\Omega^{(\prime)}_{QQ^{(\prime)}} \Omega^{(\prime)}_{QQ^{(\prime)}} \eta (\eta^\prime)}  $ 
		as  functions of continuum threshold $s_0$ at average value of $\cos\theta$.}
\label{fig:s0}
\end{figure} 

\begin{figure}[h!]
	\includegraphics[width=1.0\textwidth]{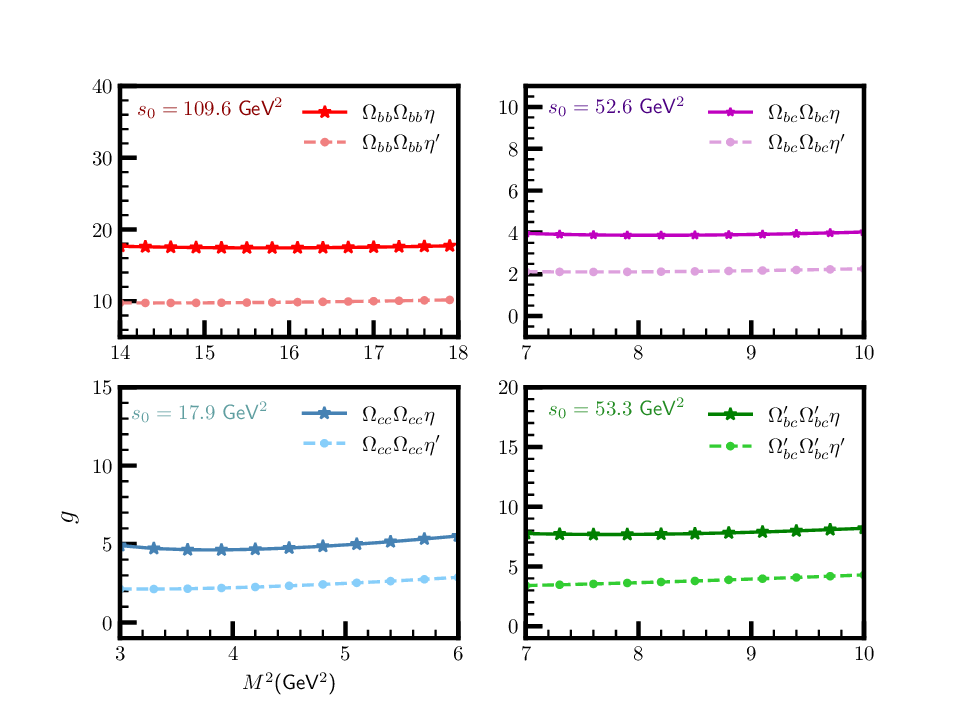}
	\caption{The strong couplings $g_{\Omega^{(\prime)}_{QQ^{(\prime)}} \Omega^{(\prime)}_{QQ^{(\prime)}} \eta (\eta^\prime)}  $ 
		as  functions of the $M^2$ at the average value of $\cos\theta$.}
	\label{fig:msq}
\end{figure}
For $M^2$, the lower and higher limits are determined as follows. The lower bound, $M^2_{\text{min}}$, is found bydemanding the OPE convergence. The higher bound, $M^2_{\text{max}}$, is determined by the requirement of pole dominance, i.e.,
\begin{eqnarray}
R=\dfrac{\int_{(m_Q+m_{Q^{\prime}})^2}^{s_0}ds \rho(s) e^{-s/M^2}}{\int_{(m_Q+m_{Q^{\prime}})^2}^{\infty}ds \rho(s) e^{-s/M^2}}\geq \frac{1}{2}.
\end{eqnarray} 
\begin{table}[h!]
	\begin{center}
		\rowcolors{2}{cyan!10}{white}
		\begin{tabular}{ c c c c }
			\rowcolor{cyan!30}
			Channel & $ M^2 $(GeV$ ^2 $) & $ s_0 $ (GeV$ ^2 $) & strong coupling constant\\
									\hline
			\hline
			\multicolumn{4}{c}{  \textcolor{cyan}{Decays to $ \pi $}} \\
			\hline
$ \Xi_{bb}\rightarrow \Xi_{bb} \pi^0  $&$ 14\leq M^2\leq18 $ &$ 105.3\leq s_0\leq113.6 $ &$  17.63^{\:2.74}_{\:2.64}  $\\
$ \Xi_{bb}\rightarrow \Xi_{bb} \pi^\pm  $&$ 14\leq M^2\leq18 $ &$ 105.3\leq s_0\leq113.6 $ &$  24.93^{\:3.87}_{\:3.73}  $\\
\hline 
$ \Xi_{bc}\rightarrow \Xi_{bc} \pi^0  $&$ 7\leq M^2\leq10 $ &$ 49.3\leq s_0\leq55$ &$3.76^{\:0.64}_{\:0.59}   $\\
$ \Xi_{bc}\rightarrow \Xi_{bc} \pi^\pm  $&$ 7\leq M^2\leq10 $ &$ 49.3\leq s_0\leq55$ &$5.32^{\:0.91}_{\:0.84}   $ \\
\hline
$ \Xi_{cc}\rightarrow \Xi_{cc} \pi^0  $ & $ 3\leq M^2\leq6 $ &$ 15.4\leq s_0\leq18.7 $ & $  5.27^{\:1.40}_{\:1.22} $ \\
$ \Xi_{cc}\rightarrow \Xi_{cc} \pi^\pm  $ & $ 3\leq M^2\leq6 $ &$ 15.4\leq s_0\leq18.7 $ & $  7.45^{\:1.98}_{\:1.71} $\\
\hline
$ \Xi^\prime_{bc}\rightarrow \Xi^\prime_{bc} \pi^0  $&$ 7\leq M^2\leq10 $ &$ 50.3\leq s_0\leq56.1 $ &$  7.84^{\:1.26}_{\:1.26}$   \\
$ \Xi^\prime_{bc}\rightarrow \Xi^\prime_{bc} \pi^\pm $&$ 7\leq M^2\leq10 $ &$ 50.3\leq s_0\leq56.1 $ &$  11.08^{\:1.78}_{\:1.79}$ \\
\hline
$ \Xi^\prime_{bc}\rightarrow \Xi_{bc} \pi^0  $&$ 7\leq M^2\leq10 $ &$ 50.3\leq s_0\leq56.1 $ &$  0.62^{\:0.18}_{\:0.17}$   \\
$ \Xi^\prime_{bc}\rightarrow \Xi_{bc} \pi^\pm $&$ 7\leq M^2\leq10 $ &$ 50.3\leq s_0\leq56.1 $ &$  0.89^{\:0.26}_{\:0.25}$ \\
\hline
			\hline
			\multicolumn{4}{c}{ \textcolor{cyan}{Decays to $K$}} \\
			\hline
			$ \Omega_{bb}\rightarrow \Xi_{bb} \bar{K}^0  $&$ 14\leq M^2\leq18 $ &$ 105.3\leq s_0\leq113.6 $ &$ 22.36^{\:4.03}_{\:3.77}  $\\
			$ \Omega_{bb}\rightarrow \Xi_{bb} K^-  $& $ 14\leq M^2\leq18 $ &$105.5\leq s_0\leq113.8 $ &$ 22.90^{\:4.12}_{\:3.80} $\\
			\hline

			$ \Omega_{bc}\rightarrow \Xi_{bc} \bar{K}^0  $&$ 7\leq M^2\leq10 $ &$ 49.7\leq s_0\leq55.5 $ &$ 4.04^{\:0.85}_{\:0.73}  $\\
			$ \Omega_{bc}\rightarrow \Xi_{bc} K^-  $& $ 7\leq M^2\leq10 $ &$49.7\leq s_0\leq55.5 $ &$  4.05^{\:0.85}_{\:0.74} $\\
			\hline

			$ \Omega_{cc}\rightarrow \Xi_{cc} \bar{K}^0  $&$ 3\leq M^2\leq6 $ &$ 16.2\leq s_0\leq19.6 $ &$  5.76^{\:1.76}_{\:1.35} $\\
			$ \Omega_{cc}\rightarrow \Xi_{cc} K^-  $& $ 3\leq M^2\leq6 $ &$ 16.2\leq s_0\leq19.6 $ &$  5.78^{\:1.78}_{\:1.38}  $\\
			\hline
			
			$ \Omega^\prime_{bc}\rightarrow \Xi^\prime_{bc} \bar{K}^0  $&$ 7\leq M^2\leq10$ &$ 50.4\leq s_0\leq56.2 $ &$  11.11^{\:2.38}_{\:2.08}  $\\
			$ \Omega^\prime_{bc}\rightarrow \Xi^\prime_{bc} K^-  $& $ 7\leq M^2\leq10 $ &$ 50.4\leq s_0\leq56.2 $ &$  11.14^{\:2.39}_{\:2.08} $\\

			\hline
			\hline
			\multicolumn{4}{c}{  \textcolor{cyan}{Decays to $\eta$}} \\
			\hline
			$ \Omega_{bb}\rightarrow \Omega_{bb} \eta  $&$ 14\leq M^2\leq18 $ &$ 105.3\leq s_0\leq113.6 $ &$ 17.20^{\:2.93}_{\:2.93}  $\\
			\hline
			$ \Omega_{bc}\rightarrow \Omega_{bc} \eta $& $ 7\leq M^2\leq10 $ &$49.7\leq s_0\leq55.5 $ &$ 3.36^{\:0.64}_{\:0.57}  $\\
			\hline
			$ \Omega_{cc}\rightarrow \Omega_{cc} \eta $&$ 3\leq M^2\leq6 $ &$ 16.2\leq s_0\leq19.6 $ &$ 4.14^{\:1.19}_{\:0.91}  $\\
			\hline
			$ \Omega^\prime_{bc}\rightarrow \Omega^\prime_{bc} \eta$& $ 7\leq M^2\leq10 $ &$ 50.4\leq s_0\leq56.2 $ &$ 8.38^{\:1.61}_{\:1.46}  $\\
			\hline
			\hline
			\multicolumn{4}{c}{  \textcolor{cyan}{Decays to $\eta^{\prime}$}} \\
			\hline
			$ \Omega_{bb}\rightarrow \Omega_{bb} \eta^\prime  $&$ 14\leq M^2\leq18 $ &$ 105.3\leq s_0\leq113.6 $ &$ 9.54^{\:1.95}_{\:1.86}  $\\
			\hline
			$ \Omega_{bc}\rightarrow \Omega_{bc} \eta^\prime $& $ 7\leq M^2\leq10 $ &$49.7\leq s_0\leq55.5 $ &$ 1.78^{\:0.41}_{\:0.37}  $\\
			\hline
			$ \Omega_{cc}\rightarrow \Omega_{cc} \eta^\prime  $&$ 3\leq M^2\leq6 $ &$ 16.2\leq s_0\leq19.6 $ &$ 1.80^{\:0.80}_{\:0.80}  $\\
			\hline
			$ \Omega^\prime_{bc}\rightarrow \Omega^\prime_{bc} \eta^\prime $& $ 7\leq M^2\leq10 $ &$ 50.4\leq s_0\leq56.2 $ &$ 4.43^{\:1.14}_{\:1.04}  $\\
			
			\hline
			
		\end{tabular}
	\end{center}
	\caption{Working regions of the Borel mass $ M^2 $ and   continuum threshold $ s_0 $, with numerical values for different strong coupling constants extracted from the analysis.} \label{tab:g} 
\end{table}

 Fig. \ref{fig:msq}  displays  the dependence of   $g_{\Omega^{(\prime)}_{QQ^{(\prime)}} \Omega^{(\prime)}_{QQ^{(\prime)}} \eta (\eta^\prime)}  $ on $M^2$ in its working window and at average values of other auxiliary parameters. From this figure we see mild variations of the results with respect to the changes in the Borel parameter $M^2$. Extracted from the analysis, the working intervals for all auxiliary parameters in all strong decay channels are depicted in Table \ref{tab:g}.

The final results for the strong coupling constants under study are also  presented in Table \ref{tab:g}. The errors in the presented results are due to the uncertainties in determinations of the working intervals for the auxiliary parameters, the intrinsic uncertainties of the method,  the errors in the masses and residues of the doubly heavy baryons, and the uncertainties coming from the DA parameters as well as other inputs. As we previously said, in Ref. \cite{Olamaei:2020bvw} we investigated the symmetric $ \Xi_ {QQ^{(\prime)}}$ couplings to the $ \pi $ meson, in which the continuum subtraction procedure is different than that of the present study. We extracted those coupling constants in the present study, as well. Comparing the results on $ g_{\Xi_{bb}\Xi_{bb}\pi^0} $, $ g_{\Xi_{bc}\Xi_{bc}\pi^0} $ and $ g_{\Xi_{cc}\Xi_{cc}\pi^0} $ from the present study and Ref. \cite{Olamaei:2020bvw}, we see that the extracted values are close to each other within the presented errors. The small differences are due to the fact that, the values in   Ref. \cite{Olamaei:2020bvw} were extracted in a single value for $ \cos\theta $ inside its working window, while in the present study we take the average of many values obtained at different values of $ \cos\theta $  in its working interval. The errors in the present study are small compared to the results of  Ref. \cite{Olamaei:2020bvw}. From Table \ref{tab:g}, it is clear that, overall, the couplings for each symmetric/anti-symmetric case and  pseudoscalar meson in the $ bb $ channels are greater than those in the $ cc $ channels, and the latter are greater than the couplings in the $ bc $ channels. In extracting the results at $ \eta $ and  $ \eta^\prime $, we have ignored  the mixing between these two states. Our results may be checked via different phenomenological approaches.

\section{Summary and concluding notes}\label{SC}
Motivated by the LHCb observation of  the $\Xi_{cc}^{++} (ccu)$ state, we have investigated the strong vertices of the doubly heavy baryons of various quark contents with the light pseudoscalar  $ \pi $, $ K $,  $\eta$ and $ \eta^\prime $ mesons. We extracted the strong coupling constants at $ q^2=m^2_{\cal P} $ from the strong coupling form factors. In the calculations, we used the general forms of the interpolating currents in their symmetric and anti-symmetric forms. We also used the light cone DAs of the pseudoscalar mesons entering the calculations. The strong coupling constants are fundamental objects and their investigation can help us get useful knowledge on the nature of the strong interactions among the participating particles. Such objects can also help us in our understanding of QCD as the theory of strong interactions. One of the main problems in studying the strong interactions between hadrons is to determine their interaction potential. The obtained results may be used in the  construction of such strong potentials.  
The obtained results may also help experimental groups in  analysis of the results obtained at hadron colliders. Investigation of doubly heavy baryons may help experiments in the course of searches for  doubly heavy baryons. Note again that, we have only one state,  $\Xi_{cc}$, has been detected  experimentally and listed in the PDG. However, even for this particle there is a tension between the SELEX and LHCb results on the mass and width of this particle. More theoretical and experimental studies  on the properties of doubly heavy baryons and their various interactions with other hadrons are needed.

\section*{Acknowledgment}

 K. Azizi and S.~Rostami are thankful  to the  Iran Science Elites Federation (Saramadan) for  financial support provided under  grant number ISEF/M/99171.

\section*{Note added:}
After completing this study and in the final stages of  proof reading we noticed that the paper Ref. \cite{Alrebdi:2020rev} appeared in arXiv, where the authors calculate the strong coupling constants among some of  the doubly heavy baryons and  $\pi$ and $K$ mesons. 

\appendix
\section{Non-local matrix elements} \label{APA}
In this Appendix, we present explicit expressions for the non-local matrix elements up to twist 4 in terms of the DAs that enter our calculations. These expressions are well known and have been taken from Refs.~ \cite{Ball:1998je,Ball:2004ye}. Here the $\pi$ meson represents all pseudoscalar mesons.
\begin{eqnarray}
\langle {\pi}(p)| \bar q(x) \gamma_\mu \gamma_5 q(0)| 0 \rangle &=& -i f_{\pi} p_\mu  \int_0^1 du  e^{i \bar u p x} 
\left( \varphi_{\pi}(u) + \frac{1}{16} m_{\pi}^2 x^2 {\mathbb A}(u) \right)
\nonumber \\
&-& \frac{i}{2} f_{\pi} m_{\pi}^2 \frac{x_\mu}{px} \int_0^1 du e^{i \bar u px} {\mathbb B}(u), 
\nonumber \\
\langle {\pi}(p)| \bar q(x) i \gamma_5 q(0)| 0 \rangle &=& \mu_{\pi} \int_0^1 du e^{i \bar u px} \varphi_P(u),
\nonumber \\
\langle {\pi}(p)| \bar q(x) \sigma_{\alpha \beta} \gamma_5 q(0)| 0 \rangle &=& 
\frac{i}{6} \mu_{\pi} \left( 1 - \tilde \mu_{\pi}^2 \right) \left( p_\alpha x_\beta - p_\beta x_\alpha\right)
\int_0^1 du e^{i \bar u px} \varphi_\sigma(u),
\nonumber \\
\langle {\pi}(p)| \bar q(x) \sigma_{\mu \nu} \gamma_5 g_s G_{\alpha \beta}(v x) q(0)| 0 \rangle &=&
i \mu_{\pi} \left[
p_\alpha p_\mu \left( g_{\nu \beta} - \frac{1}{px}(p_\nu x_\beta + p_\beta x_\nu) \right) 
\right. \nonumber \\
&-&	p_\alpha p_\nu \left( g_{\mu \beta} - \frac{1}{px}(p_\mu x_\beta + p_\beta x_\mu) \right) 
\nonumber \\
&-&	p_\beta p_\mu \left( g_{\nu \alpha} - \frac{1}{px}(p_\nu x_\alpha + p_\alpha x_\nu) \right)
\nonumber \\ 
&+&	p_\beta p_\nu \left. \left( g_{\mu \alpha} - \frac{1}{px}(p_\mu x_\alpha + p_\alpha x_\mu) \right)
\right]
\nonumber \\
&\times& \int {\cal D} \alpha e^{i (\alpha_{\bar q} + v \alpha_g) px} {\cal T}(\alpha_i),
\nonumber \\
\langle {\pi}(p)| \bar q(x) \gamma_\mu \gamma_5 g_s G_{\alpha \beta} (v x) q(0)| 0 \rangle &=& 
p_\mu (p_\alpha x_\beta - p_\beta x_\alpha) \frac{1}{px} f_{\pi} m_{\pi}^2 
\int {\cal D}\alpha e^{i (\alpha_{\bar q} + v \alpha_g) px} {\cal A}_\parallel (\alpha_i)
\nonumber \\
&+& \left[
p_\beta \left( g_{\mu \alpha} - \frac{1}{px}(p_\mu x_\alpha + p_\alpha x_\mu) \right) \right.
\nonumber \\
&-& 	p_\alpha \left. \left(g_{\mu \beta}  - \frac{1}{px}(p_\mu x_\beta + p_\beta x_\mu) \right) \right]
f_{\pi} m_{\pi}^2
\nonumber \\
&\times& \int {\cal D}\alpha e^{i (\alpha_{\bar q} + v \alpha _g) p x} {\cal A}_\perp(\alpha_i),
\nonumber \\
\langle {\pi}(p)| \bar q(x) \gamma_\mu i g_s G_{\alpha \beta} (v x) q(0)| 0 \rangle &=& 
p_\mu (p_\alpha x_\beta - p_\beta x_\alpha) \frac{1}{px} f_{\pi} m_{\pi}^2 
\int {\cal D}\alpha e^{i (\alpha_{\bar q} + v \alpha_g) px} {\cal V}_\parallel (\alpha_i)
\nonumber \\
&+& \left[
p_\beta \left( g_{\mu \alpha} - \frac{1}{px}(p_\mu x_\alpha + p_\alpha x_\mu) \right) \right.
\nonumber \\
&-& 	p_\alpha \left. \left(g_{\mu \beta}  - \frac{1}{px}(p_\mu x_\beta + p_\beta x_\mu) \right) \right]
f_{\pi} m_{\pi}^2
\nonumber \\
&\times& \int {\cal D}\alpha e^{i (\alpha_{\bar q} + v \alpha _g) p x} {\cal V}_\perp(\alpha_i).
\end{eqnarray}

\section{DAs for the light pseudoscalar mesons} \label{DAs}

Here we present the DAs for the light pseudoscalar mesons which  appear in our calculations. They are well known and taken from Refs. \cite{Ball:1998je,Ball:2004ye}.

\begin{eqnarray}
\phi_{\pi}(u) &=& 6 u \bar u \Big( 1 + a_1^{\pi} C_1(2 u -1) + a_2^{\pi} C_2^{3 \over 2}(2 u - 1) \Big), 
\nonumber \\
{\cal T}(\alpha_i) &=& 360 \eta_3 \alpha_{\bar q} \alpha_q \alpha_g^2 \Big( 1 + w_3 \frac12 (7 \alpha_g-3) \Big),
\nonumber \\
\phi_P(u) &=& 1 + \Big( 30 \eta_3 - \frac{5}{2} \frac{1}{\mu_{\pi}^2}\Big) C_2^{1 \over 2}(2 u - 1) 
\nonumber \\ 
&+&	\Big( -3 \eta_3 w_3  - \frac{27}{20} \frac{1}{\mu_{\pi}^2} - \frac{81}{10} \frac{1}{\mu_{\pi}^2} a_2^{\pi} \Big) C_4^{1\over2}(2u-1),
\nonumber \\
\phi_\sigma(u) &=& 6 u \bar u \Big[ 1 + \Big(5 \eta_3 - \frac12 \eta_3 w_3 - \frac{7}{20}  \mu_{\pi}^2 - \frac{3}{5} \mu_{\pi}^2 a_2^{\pi} \Big)
C_2^{3\over2}(2u-1) \Big],
\nonumber \\
{\cal V}_\parallel(\alpha_i) &=& 120 \alpha_q \alpha_{\bar q} \alpha_g \Big( v_{00} + v_{10} (3 \alpha_g -1) \Big),
\nonumber \\
{\cal A}_\parallel(\alpha_i) &=& 120 \alpha_q \alpha_{\bar q} \alpha_g \Big( 0 + a_{10} (\alpha_q - \alpha_{\bar q})\Big),
\nonumber \\
{\cal V}_\perp (\alpha_i) &=& - 30 \alpha_g^2\Big[ h_{00}(1-\alpha_g) + h_{01} (\alpha_g(1-\alpha_g)- 6 \alpha_q \alpha_{\bar q}) +
h_{10}(\alpha_g(1-\alpha_g) - \frac32 (\alpha_{\bar q}^2+ \alpha_q^2)) \Big],
\nonumber \\
{\cal A}_\perp (\alpha_i) &=& 30 \alpha_g^2(\alpha_{\bar q} - \alpha_q) \Big[ h_{00} + h_{01} \alpha_g + \frac12 h_{10}(5 \alpha_g-3) \Big],
\nonumber \\
B(u)&=& g_{\pi}(u) - \phi_{\pi}(u),
\nonumber \\
g_{\pi}(u) &=& g_0 C_0^{\frac12}(2 u - 1) + g_2 C_2^{\frac12}(2 u - 1) + g_4 C_4^{\frac12}(2 u - 1),
\nonumber \\
{\mathbb A}(u) &=& 6 u \bar u \left[\frac{16}{15} + \frac{24}{35} a_2^{\pi}+ 20 \eta_3 + \frac{20}{9} \eta_4 +
\Big( - \frac{1}{15}+ \frac{1}{16}- \frac{7}{27}\eta_3 w_3 - \frac{10}{27} \eta_4 \right) C_2^{3 \over 2}(2 u - 1) 
\nonumber \\ 
&+&\Big( - \frac{11}{210}a_2^{\pi} - \frac{4}{135} \eta_3w_3 \Big)C_4^{3 \over 2}(2 u - 1)\Big]
\nonumber \\
&+& \Big( -\frac{18}{5} a_2^{\pi} + 21 \eta_4 w_4 \Big)\Big[ 2 u^3 (10 - 15 u + 6 u^2) \ln u 
\nonumber \\
&+& 2 \bar u^3 (10 - 15 \bar u + 6 \bar u ^2) \ln\bar u + u \bar u (2 + 13 u \bar u) \Big],
\end{eqnarray}
where $C_n^k(x)$ are the Gegenbauer polynomials and 
\begin{eqnarray}
h_{00}&=& v_{00} = - \frac13\eta_4,
\nonumber \\
a_{10} &=& \frac{21}{8} \eta_4 w_4 - \frac{9}{20} a_2^{\pi},
\nonumber \\
v_{10} &=& \frac{21}{8} \eta_4 w_4,
\nonumber \\
h_{01} &=& \frac74  \eta_4 w_4  - \frac{3}{20} a_2^{\pi},
\nonumber \\
h_{10} &=& \frac74 \eta_4 w_4 + \frac{3}{20} a_2^{\pi},
\nonumber \\
g_0 &=& 1,
\nonumber \\
g_2 &=& 1 + \frac{18}{7} a_2^{\pi} + 60 \eta_3  + \frac{20}{3} \eta_4,
\nonumber \\
g_4 &=&  - \frac{9}{28} a_2^{\pi} - 6 \eta_3 w_3.
\label{param0}
\end{eqnarray}
The constants inside the wavefunctions which are calculated at
the renormalization scale of $\mu=1 ~\mbox{GeV}^{2}$ are given as
$a_{1}^{\pi} = 0$, $a_{2}^{\pi} = 0.44$,
$\eta_{3} =0.015$, $\eta_{4}=10$, $w_{3} = -3$ and 
$ w_{4}= 0.2$.

\end{document}